\documentclass[%
 reprint,superscriptaddress,
 amsmath,amssymb,
 aps,prl
]{revtex4-1}
\usepackage{graphicx}
\usepackage{dcolumn}
\usepackage{bm}
\usepackage[caption=false]{subfig}
\usepackage{floatrow}
\usepackage{appendix}
\usepackage{amsmath}
\usepackage[italicdiff]{physics}
\usepackage{hyperref}
\usepackage[capitalise]{cleveref}
\usepackage{xcolor}

\DeclareCaptionLabelFormat{adja-page}{\hrulefill\\#1 #2 \emph{(previous page)}}

\begin{abstract}
Whether or not anomalies in the thermal conductivity in insulating cuprates can be attributed to antiferromagnetic order and magnons in a 2D Mott insulator remains an intriguing open question. To shed light on this issue, we investigate the thermal conductivity $\kappa$ and its relationship with the specific heat $c_v$ in the half-filled 2D single-band Hubbard model, using the numerically exact determinant quantum Monte Carlo algorithm and maximum entropy analytic continuation.
At low temperatures where the charge degrees of freedom are gapped-out and $c_v$ exhibits a clear magnon peak, 
we observe that thermal conductivity $\kappa$ also tends to form a peak at similar temperatures. Reducing temperature further produces a sharp upturn in $\kappa$, associated with an increasing mean-free path. We identify this as the high-temperature side of the anomalous peak in insulating cuprates, where the mean-free path eventually is cut-off by other scattering effects, including phonons, disorder, and physical size.
Different scattering effects in our model are identified and analyzed in the thermal diffusivity.

\end{abstract}

\begin{document}

\title{Magnon heat transport in a two-dimensional Mott insulator}

\author{Wen O. Wang}
\email{wenwang.physics@gmail.com}
\affiliation{Department of Applied Physics, Stanford University, Stanford, California 94305, USA}

\author{Jixun K. Ding}
\affiliation{Department of Applied Physics, Stanford University, Stanford, California 94305, USA}

\author{Brian Moritz}
\affiliation{Stanford Institute for Materials and Energy Sciences,
SLAC National Accelerator Laboratory, 2575 Sand Hill Road, Menlo Park, California 94025, USA}

\author{Edwin W. Huang}
\affiliation{Department of Physics and Institute of Condensed Matter Theory, University of Illinois at Urbana-Champaign, Urbana, Illinois 61801, USA}

\author{Thomas P. Devereaux}
\email{tpd@stanford.edu}
\affiliation{Stanford Institute for Materials and Energy Sciences,
SLAC National Accelerator Laboratory, 2575 Sand Hill Road, Menlo Park, California 94025, USA}
\affiliation{
Department of Materials Science and Engineering, Stanford University, Stanford, California 94305, USA}
\date{\today}

\maketitle

\section{Introduction}

The effects of magnetic ordering on transport in the high-$T_c$ cuprates are topics of great interest.
For undoped strongly correlated systems that are electrically insulating due to Mott physics,
heat transport can be measured to probe the excitations~\cite{PhysRevLett.88.095901,PhysRevLett.91.146601,PhysRevLett.90.197002,NAKAMURA19911409,PhysRevLett.95.156603}, in analogy to how charge transport probes excitations in the metallic phase.
For a wide range of insulating antiferromagnetic cuprates, a general two-peak structure appears in the temperature dependence of thermal conductivity.
A low-temperature phonon-related peak at $\sim 25 \, \mathrm{K}$ is present in both in-plane and out-of-plane thermal conductivity, and an additional anomalously broad peak at temperature $\sim 250 \, \mathrm{K}$ has been observed in the in-plane thermal conductivity~\cite{PhysRevB.73.104430,PhysRevLett.90.197002,PhysRevB.67.104503,NAKAMURA19911409,PhysRevB.67.184502}.
While considerable experimental evidence suggests that this high temperature anomaly arises from magnons or magnetic excitations~\cite{PhysRevB.67.184502,PhysRevB.73.104430,PhysRevLett.90.197002,PhysRevB.92.054409}, its origin remains unclear
~\cite{PhysRevB.39.804,PhysRevB.52.R13134}.
Resolving this debate about the origin of the anomalous peak and understanding its microscopic dynamics requires 
further calculations of magnon contributions to heat transport in such systems.

The calculation of transport properties in strongly correlated many-body systems presents a formidable challenge.
For an antiferromagnetic Mott insulator,
a typical theoretical description for thermal transport~\cite{Shastry_2008,PhysRev.135.A1505} begins by applying
spin-wave theory~\cite{PhysRev.87.568} to an antiferromagnetic Heisenberg model, which leads to low-energy dispersive magnetic excitations, i.e., magnons.
Boltzmann theory can then be applied, assuming that magnons are well defined and weakly interacting~\cite{dicastro_raimondi_2015, ashcroft1976solid,PhysRevLett.87.047202,PhysRevB.67.184502}.
However, it is hard to verify whether or at which temperatures these assumptions are correct.
Attempts to study thermal transport of magnons often involve taking various limits~\cite{PhysRevB.39.2344,PhysRevB.41.9228,PhysRevB.42.2096,PhysRevLett.89.156603,PhysRevB.66.140406,PhysRevLett.92.069703}.
Moreover, even if we assume Boltzmann theory is valid,
heat transport remains difficult to calculate, as precise information about magnon scattering is lacking. 
Exact calculations for magnon heat transport without simplifying assumptions has been an extreme challenge that has remained relatively unexplored for strongly correlated systems. 

The Hubbard model has been widely studied as a simplified description of the electronic properties of high-$T_c$ cuprates~\cite{RevModPhys.87.457,Keimer2015}.
Although the model lacks an analytic solution in two dimensions (2D),
several unconventional transport phenomena in cuprates are successfully captured in numerical simulations~\cite{Huang987, Wang2020, PhysRevResearch.3.033033} and in cold atom experiments~\cite{Xu2019,Brown379,Nichols383}.  
The determinant quantum Monte Carlo (DQMC) algorithm~\cite{dqmc1,dqmc2} and maximum entropy analytic continuation (MaxEnt)~\cite{ana1,ana2} have recently been utilized to investigate optical conductivity, successfully finding strange metallicity in the doped model and insulating behavior at half-filling~\cite{Huang987}.

\begin{figure}
    \centering
    \includegraphics[width=\textwidth]{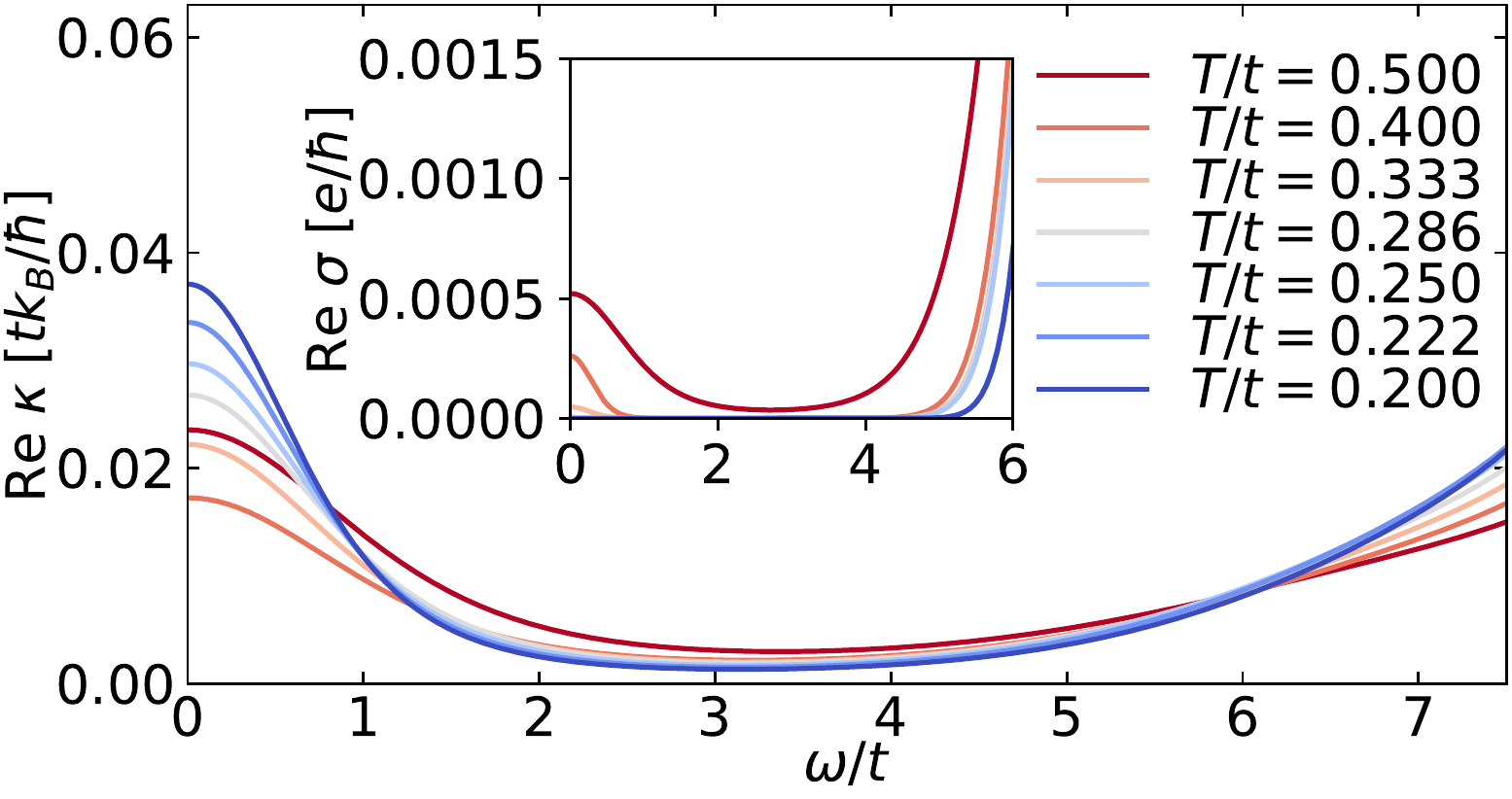}
    \caption{Frequency dependence of the real parts of thermal conductivity $\Re \kappa(\omega)$ and optical conductivity $\Re \sigma(\omega)$ (inset)~\cite{Huang987} for the half-filled Hubbard model with $U/t=12$ at different temperatures.
    Simulation lattice size is $8 \times 8$.
    }
    \label{fig:frequency}
\end{figure}

In the limit of strong correlations $t/U \ll 1$ and low temperatures $T/U \ll 1$, projecting out doubly occupied states in the half-filled Hubbard model produces an effective spin-$\frac{1}{2}$ %\frac{1}{2}$ 
antiferromagnetic Heisenberg low-energy model~\cite{Auerbach1994heisenberg}, with spin exchange energy $J=4t^2/U$.
Here, $t$ is the nearest-neighbor hopping energy, $U$ is the Coulomb interaction, and $T$ is the temperature.
At strong coupling, a nonzero-temperature maximum of the local moment $\ev{m_z^2}$~\cite{PhysRevB.63.125116} and a sharp peak of the spin-spin correlator $S(\mathbf{q})$ at $\mathbf{q}=(\pi,\pi)$ in DQMC~\cite{PhysRevB.63.125116,PhysRevB.80.075116,PhysRevB.93.155166,dqmc2} convincingly demonstrate the formation of antiferromagnetic magnons at temperature scales below $J$.
Using MaxEnt, %~\cite{ana1,ana2}, 
the dynamical spin structure factor $S(\mathbf{q},\omega)$ has been calculated~\cite{PhysRevB.92.195108,Huang1161,Huang2018} for the undoped Hubbard model, where the results agree with spin-wave theory.

DQMC is a numerically exact algorithm, especially efficient for calculations of the half-filled Hubbard model, and when including only nearest-neighbor hopping, the model preserves particle-hole symmetry and is sign-problem free~\cite{SIGN}, enabling simulations on large lattices down to low temperatures.
We are thus motivated to use DQMC~\cite{dqmc1,dqmc2} and MaxEnt~\cite{ana1,ana2,PhysRevE.94.023303} to investigate thermal transport properties  of the half-filled Hubbard model, particularly when antiferromagnetic correlations are strong~\cite{PhysRevB.35.3359,PhysRevB.38.316,PhysRevB.39.2344}. 
In contrast to optical conductivity and spin dynamical response, which involve four fermion operators, thermal conductivity requires measuring heat current-heat current correlation functions, which involve observables with up to eight fermion operators, indicating that a greater amount of simulation data is required to obtain converged and accurate results. This has been one of the key challenges that has precluded an analysis of thermal conductivity in the Hubbard model in the past. In the Supplementary Material~\cite{supplementary}, we discuss the methodology and challenges of this calculation, including an analysis of Trotter~\cite{PhysRevB.36.3833} and finite-size errors.

In this paper, we report magnon heat transport in both the frequency and temperature dependence of the thermal conductivity for the undoped single-band 2D Hubbard model with only nearest-neighbor hopping.
We observe a Drude peak in the real part of thermal conductivity $\Re \kappa(\omega)$ at temperatures below the spin exchange energy $J$.
In the temperature dependence of the DC thermal conductivity $\kappa$,
we observe peaks at $T\sim U$ and $T\sim J$ concurrent with features in the specific heat $c_v$, and an additional sharp upturn as $T$ further decreases.
The interaction and temperature dependence are analyzed by comparing the results with the $t=0$ single-site Hubbard model at high temperatures and to the Heisenberg model at low temperatures.
We identify two contributions to $\kappa$ and $c_v$: one which involves the local kinetic energy and another which involves the interactions.
Different scattering effects are identified and analyzed in the thermal diffusivity $D_Q$.
We conclude with a comparison of the upturn of $\kappa$ with experimental results for undoped cuprates.
We leave a discussion about the consistency between the kinetic parts of both the specific heat and the thermal Drude weight~\cite{PhysRevB.66.140406} and predictions from spin-wave theory to the Supplementary Material~\cite{supplementary}.

\section{Results}

Figure~\ref{fig:frequency} shows the temperature evolution of the frequency dependence of the real parts of thermal conductivity $\Re \kappa(\omega)$ and optical conductivity $\Re \sigma(\omega)$ (inset)~\cite{Huang987}.
As temperature $T$ decreases, low-frequency $\Re \sigma(\omega)$ shows insulating behavior, while $\Re \kappa(\omega)$ shows a Drude peak close to $\omega=0$,
indicating that the antiferromagnetic magnons carry heat but no charge.
The low-frequency Drude peak of $\Re \kappa(\omega)$ becomes sharper as temperature decreases, reflecting well-defined magnons with a  decreasing scattering rate in this temperature regime. 
Profiles over a wider energy range are shown in the Supplementary Material~\cite{supplementary}.

\begin{figure}
    \centering
    \includegraphics[width=\textwidth]{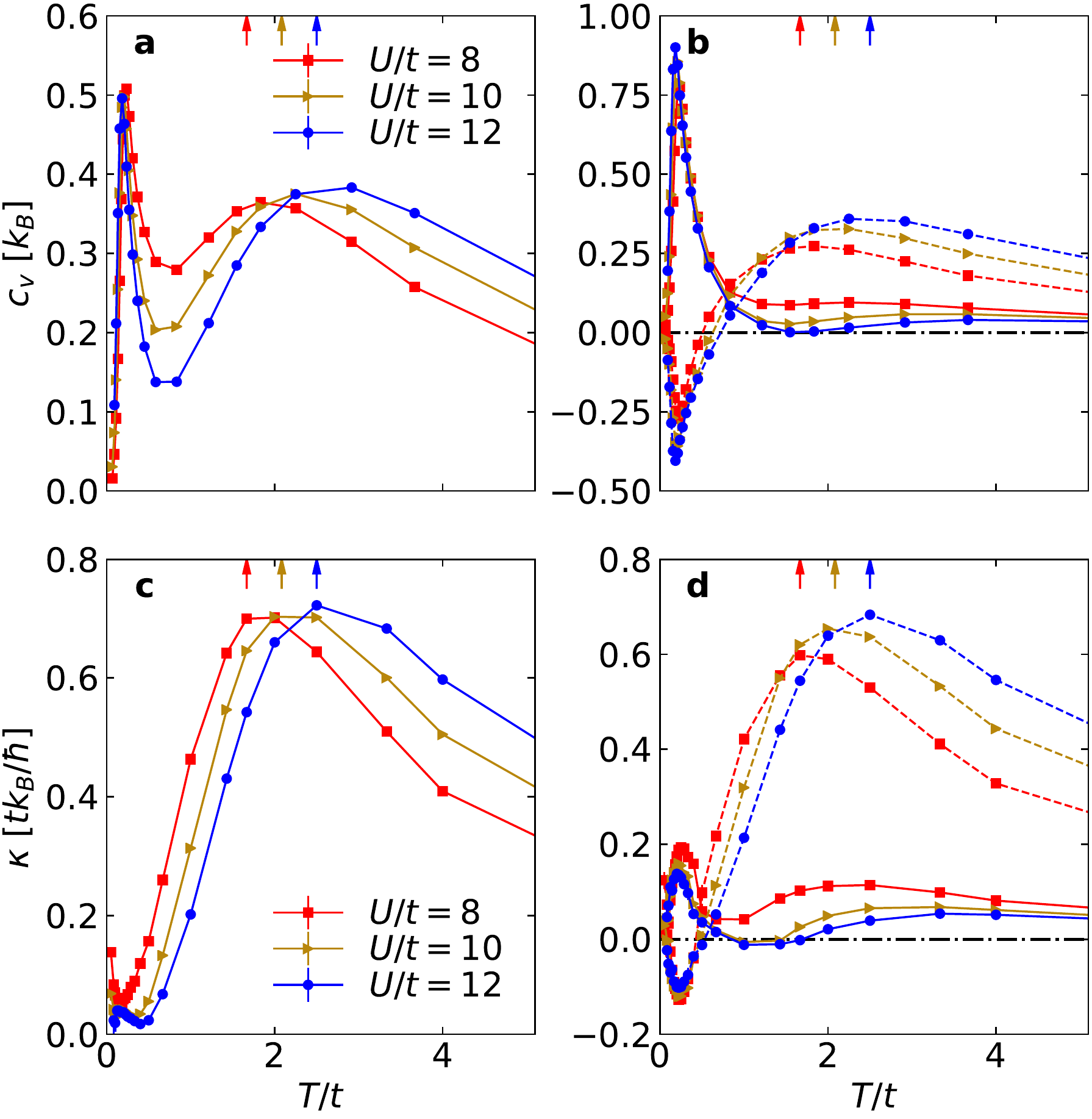}
    \caption{Specific heat $c_v$ and DC thermal conductivity $\kappa$ for the half-filled Hubbard model with $U/t=8$, $10$, and $12$.
    (\textbf{a}) The total specific heat $c_v$ calculated from finite differences. 
    (\textbf{b}) The kinetic part $c_K$ (solid lines) and potential part $c_P$ (dashed lines)  of the specific heat $c_v$ 
    calculated by finite differences.
    (\textbf{c}) The total DC thermal conductivity $\kappa$.
    (\textbf{d}) The kinetic part $\kappa_K$ (solid lines) and potential part $\kappa_P$ (dashed lines)  of $\kappa$.
    In all subplots, arrows point to $T=U/4.8$, corresponding to peak positions of the specific heat for the $t=0$ single-site Hubbard model.
    The same color (marker) represents the same $U$.
    For $c_v$, $c_K$ and $c_P$ calculated from finite differences, the error has been calculated by propagating the standard error from jackknife resampling~\cite{jackknife}, and they are smaller than the size of the data points.
    The error bars for the results for $\kappa$, $\kappa_K$ and $\kappa_P$ represent  $\pm 1$ bootstrap standard error~\cite{bootstrap}.
    Simulation lattice size is $8 \times 8$.
    }
    \label{fig:summaryhight}
\end{figure}

\begin{figure}
    \centering
    \includegraphics[width=\textwidth]{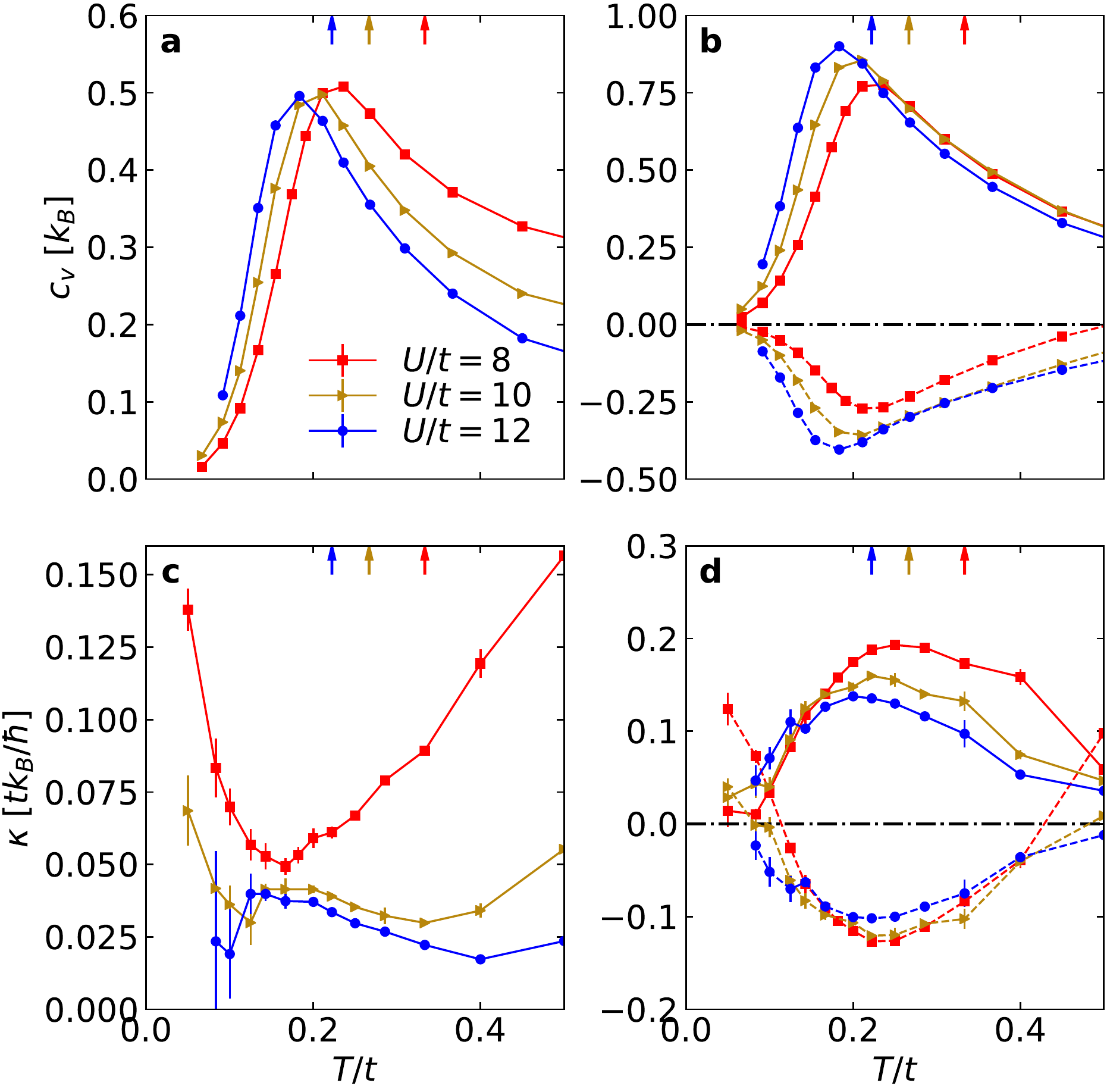}
    \caption{Same results as Fig.~\ref{fig:summaryhight}, highlighting low temperature.
    The arrows point to $T = 2J/3$, the numerically predicted peak positions for the specific heat of the corresponding Heisenberg model with $J=4t^2/U$~\cite{PhysRevLett.77.892}.
    }
    \label{fig:summarylowt}
\end{figure}

We now turn to specific heat $c_v$ and the DC limit of thermal conductivity $\kappa(\omega=0)$. %, 
For brevity, we use $\kappa$ to denote $\kappa(\omega=0)$ throughout the remainder of the paper.
Results for $c_v$ and $\kappa$ are shown in Fig.~\ref{fig:summaryhight}, with Fig.~\ref{fig:summarylowt} highlighting the low-temperature features.
For each $U$ in Fig.~\ref{fig:summaryhight}(\textbf{a}), we observe two peaks in $c_v$ which appear at two different temperatures, consistent with previous studies~\cite{PhysRevB.63.125116,PhysRevB.55.12918,PhysRevA.86.023633}: 
a low-temperature peak associated with the spin exchange energy $J$, reflecting the formation of antiferromagnetic magnons as $T$ decreases, and
a high-temperature peak associated with the Coulomb interaction $U$, reflecting the suppression of double occupancy as $T$ decreases.
The high-temperature peak positions are close to $U/4.8$~\cite{PhysRevB.63.125116}, which is the predicted peak position of the $t=0$ single-site Hubbard model.
The low-temperature peak positions
in Fig.~\ref{fig:summarylowt}(\textbf{a}) deviate from the $c_v$ peak position numerically predicted in the Heisenberg model $T\sim 2J/3$~\cite{PhysRevLett.77.892}, in contrast with previous results~\cite{PhysRevB.63.125116,PhysRevB.55.12918}, as we measure $c_v$ on a larger lattice, using a smaller imaginary time discretization $\dd\tau$.
This deviation is discussed in detail in the Supplementary Material~\cite{supplementary,PhysRevLett.97.187202,TANG2013557,PhysRevA.84.053611}.
As $U$ increases and the Heisenberg model becomes a better low-energy effective theory, the peak position approaches $\sim 2J/3$.

In the semiclassical kinetic theory, for a dilute gas, $\kappa$ is related to $c_v$ by $\kappa = c_v \ev{v} l / d$~\cite{dicastro_raimondi_2015, ashcroft1976solid}, where $\ev{v}$ is the particle velocity, $l$ is the mean-free path and $d$ is the number of dimensions.
For our strongly correlated system with temperatures ranging over various energy scales, this phenomenological relation is not directly applicable for quantitative behaviors without proper scattering information, 
but implies possible correspondence between $c_v$ and $\kappa$, which can be different for different temperature scales.
In Fig.~\ref{fig:summaryhight}(\textbf{c}), we find that the peak associated with $U$ also appears in $\kappa$. %two
Between the temperature scales set by $J$ and $U$, $\kappa$ drops quickly as the temperature $T$ decreases.
Below $T \sim J$ where the additional peak appears in $c_v$, as shown in Fig.~\ref{fig:summarylowt}(\textbf{a}), $\kappa$ increases again and also tends to form a peak, which becomes more apparent for strong interactions $U/t\gtrsim 10$ due to a better separation of energy scales between $J$ and $U$, as shown in Fig.~\ref{fig:summarylowt}(\textbf{c}).
As temperatures further decrease, $c_v$ decays toward $0$, but $\kappa$ shows an upturn and continues increasing, down to the lowest temperatures, associated with an increasing mean-free path $l$.
We identify this regime with the high-temperature side of the anomalous peak in insulating cuprates. Experimentally, $l$ will be cut off by various scattering effects, including phonons, disorder, and physical size, inevitably leading to the formation of a peak at lower temperatures.

To understand the temperature dependence and identify contributions from different energy scales, we separate out the kinetic (K) and potential contributions (P) to $c_v$ and $\kappa$~\cite{split1,split2} in Figs.~\ref{fig:summaryhight}(\textbf{b}) and (\textbf{d}), by splitting the total energy $H$ and defining the hopping energy as the kinetic energy $H_K$ and the electron-electron interaction as the potential energy $H_P$.
Details about the definitions and methods are in the Supplementary Material~\cite{supplementary}.
For both $c_v$ and $\kappa$, we see that the high-temperature peak mainly comes from the potential part $c_P$ and $\kappa_P$ respectively, as shown in Figs.~\ref{fig:summaryhight}(\textbf{b}) and (\textbf{d}), associated with suppression of on-site double occupancy as temperature decreases. 

At $T\sim J$, magnon peaks in $c_v$ and $\kappa$ mainly arise from the kinetic parts $c_K$ and $\kappa_K$, as shown in Figs.~\ref{fig:summarylowt}(\textbf{b}) and (\textbf{d}), consistent with our expectations,
since $J$ arises from virtual hopping processes~\cite{PhysRevB.63.125116}. 
The potential energy involves only double occupancy terms, and double occupancies are projected out when mapping the half-filled Hubbard model to the Heisenberg model, so the magnon peaks should not come from the potential parts $c_P$ and $\kappa_P$.
However, we note that, at the values of $U/t$ considered here, double occupancies are not fully suppressed, and hence, we find potential energy contributions to $c_v$ and $\kappa$ that are negative, with magnitudes still significant compared with the kinetic parts, as shown in  Figs.~\ref{fig:summarylowt}(\textbf{b}) and (\textbf{d}).
This negative dip for $c_P$ also is shown and discussed in Ref.~\cite{PhysRevB.63.125116}.
For the case of a nonzero $t'$, down to the lowest temperatures we can achieve, which is constrained by the fermion sign problem~\cite{SIGN} due to broken particle-hole symmetry, the behavior for $c_v$, $\kappa$, and their respective kinetic-potential separations shows no qualitative differences (see Supplementary Material~\cite{supplementary}).

Notably, as $T$ further decreases and the system approaches the antiferromagnetic ground state, 
where both $c_K$ and $c_P$ approach $0$, there is a switch in the dominant contribution to $\kappa$. The low-temperature upturn in $\kappa$, shown in Fig.~\ref{fig:summarylowt} (\textbf{c}), mainly comes from $\kappa_P$, as shown in Fig.~\ref{fig:summarylowt} (\textbf{d}). The switch to $\kappa_P$ is obvious especially for smaller $U$ and indicates different energy transport properties between high- and low-energy magnons.
For an intuitive understanding, consider the terms involved in the kinetic and potential energy current operators (see Supplementary Material~\cite{supplementary}). The kinetic energy current operators involve next-nearest and next-next-nearest-neighbor hoppings, which are forbidden by Pauli exclusion for an antiferromagnetic spin pattern,
while terms in the potential energy current are allowed at the expense of forming double occupancies, costing energy $\sim U$.

\begin{figure}
    \centering
    \includegraphics[width=\textwidth]{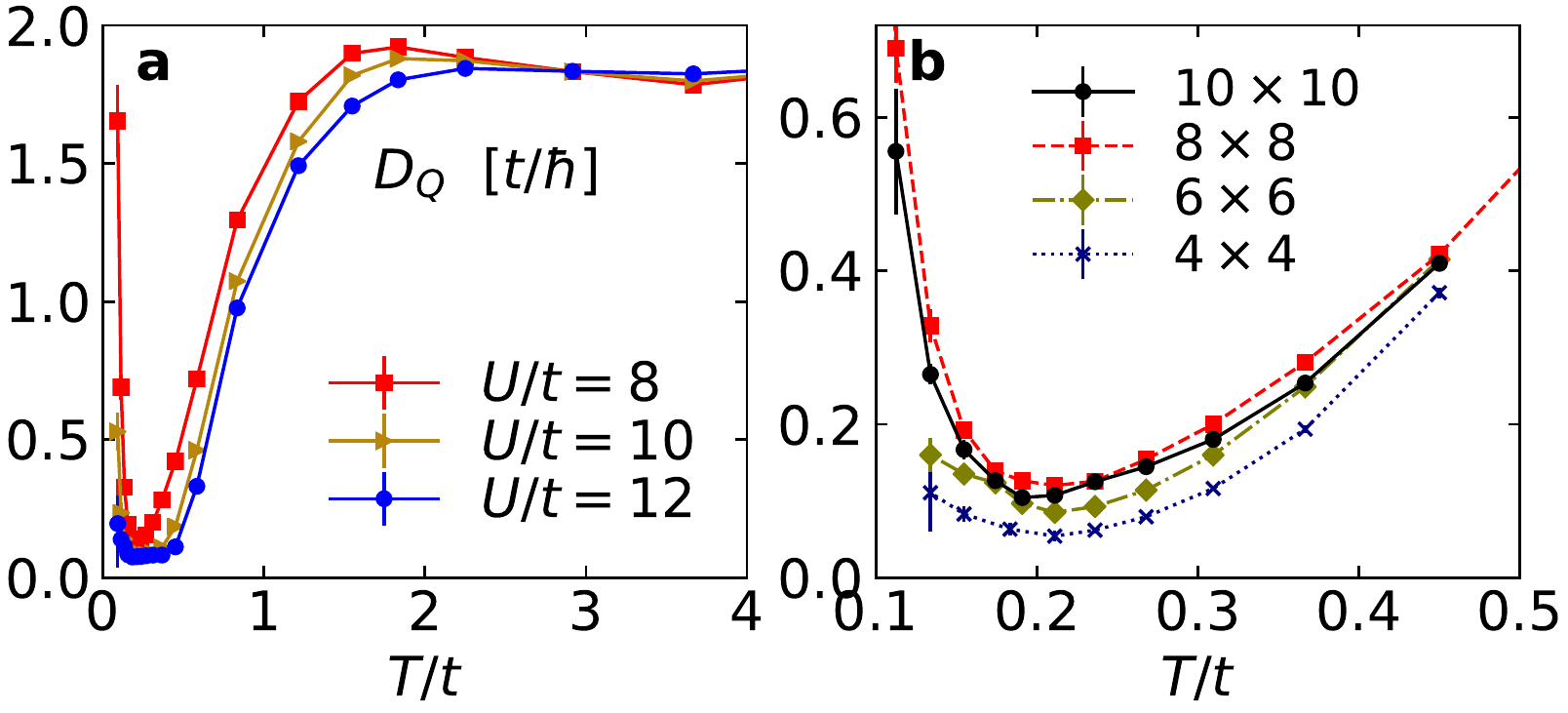}
    \caption{(\textbf{a}) Thermal diffusivity $D_Q=\kappa/c_v$ for $U/t=8$, $10$, and $12$. Simulation lattice size is $8\times 8$.
   (\textbf{b})  Temperature and lattice size dependence of $D_Q$ for $U/t=8$ at low temperatures. 
   To calculate $D_Q$ at temperature $T=(T_1+T_2)/2$, $c_v$ is calculated from finite differences of the energies at $T_1$ and $T_2$, and $\kappa$ is determined by the average of $\kappa$ obtained at temperatures $T_1$ and $T_2$.
   The error bars are calculated from error propagation assuming $c_v$ and $\kappa$ are measured independently.
    }
    \label{fig:diffusivity}
\end{figure}

Finally, to analyze the scattering mechanisms, we calculate the thermal diffusivity $D_Q = \kappa/c_v$, as shown in Fig.~\ref{fig:diffusivity}(\textbf{a}).
When temperature is high enough and the system is metallic, $D_Q$ shows weak temperature dependence for temperatures $T/t \gtrsim 1$, like the behavior of charge diffusivity $D$~\cite{Huang987}. This weak temperature dependence reflects the similarity of the temperature dependence between $\kappa$ and  $c_v$ around their high-temperature peaks.
For $T/t \lesssim 1$, we observe that $D_Q$ drops quickly as temperature decreases, signifying a switch to magnon dominated transport. 
Below this temperature scale, $D_Q$ behaves significantly differently than $D$, as opposed to the expected behavior in a Fermi liquid where the temperature dependence is similar for the two quantities.

As mentioned, according to the kinetic theory, $D_Q$ is a proxy to the phenomenological $\langle v \rangle l/d$, and thus reflects the evolution of scattering.
If one assumes magnon velocity $\ev{v}$ to be weakly temperature dependent, the temperature dependence of the mean-free path $l$ should follow $D_Q$.
In our system, possible scattering mechanisms include boundary scattering, correlation length, and magnon-magnon scattering~\cite{PhysRevB.92.054409}. Here, we discuss their respective temperature dependence trends.
The mean-free path $l$ is constrained by the lattice size and the correlation length $\xi$~\cite{PhysRevB.67.184502}, which itself increases with decreasing temperatures~\cite{PhysRevLett.104.066406,PhysRevLett.64.1449} and saturates to the order of the lattice size at some temperature (see Supplementary Material~\cite{supplementary} for behavior of the correlation length $\xi$).
The magnon-magnon scattering effects are reduced at lower %$T$ 
temperatures due to the reduced number of high-energy magnons involved in Umklapp processes for the scattering of low-energy magnons, as well as reduced scattering between the low-energy magnon branches around $\mathbf{k}=(0,0)$ and $(\pi,\pi)$~\cite{PhysRevB.92.054409}. 
These trends can be verified in our data for $D_Q$.
Figure~\ref{fig:diffusivity}(\textbf{b}) shows the temperature and lattice size dependence of $D_Q$ for $U/t=8$ at low temperatures.
For lattice sizes smaller than $8\times 8$, $D_Q$ at the lowest temperatures shows significant size dependence and tends to saturate with decreasing temperatures, indicating that $l$ is constrained by the lattice size.
Results on lattices of size $8\times 8$ and $10 \times 10$ show minimal change in $D_Q$ because $l$ is no longer constrained by the lattice size.
Here, $D_Q$ increases as $T$ decreases, reflecting larger $l$ with increasing $\xi$ and reduced magnon-magnon scattering effects.
Similar trends in $D_Q$ also are observed for $U/t=10$ and $12$ (see Supplementary Material~\cite{supplementary}).

\section{Discussion}

Here we discuss the comparison of our lowest-temperature upturn in $\kappa$ with cuprates experiments.
Our units $tk_B \hbar^{-1}$ for thermal conductivity become  $tk_B\hbar^{-1}d_z^{-1}$ for a three-dimensional material, where $d_z$ is the lattice constant perpendicular to the 2D planes.
Using $t/k_B \approx 4000\,\mathrm{K}$ and lattice constant $d_z= 13.2\,\mathring{\mathrm{A}}$, as appropriate for  La$_2$CuO$_4$~\cite{PhysRevLett.67.3622,PhysRevLett.86.5377,PhysRevB.66.144420,https://doi.org/10.1002/pssb.200301719,PhysRevLett.90.197002}, 
we find $tk_B/(d_z\hbar) \approx 5.48 \mathrm{\, W \, m^{-1}K^{-1}}$.
Thus, the experimentally reported $\kappa$ peak in La$_2$CuO$_4$ at $\sim 300\, \mathrm{K}$ has a magnitude of $\sim 2 tk_B \hbar^{-1}d_z^{-1}$~\cite{PhysRevLett.90.197002}, which is an order of magnitude higher than our results for $\kappa$ in Fig.~\ref{fig:summarylowt}(\textbf{c}).
Aside from other experimental factors that may affect the magnitude, we speculate that the most significant reason for this discrepancy is that our calculations are performed in strictly 2D, where long-range antiferromagnetic order cannot survive at finite temperature. On the other hand, La$_2$CuO$_4$ has some weak interlayer exchange coupling $J'$ between the CuO$_2$ planes, and shows a transition to long-range antiferromagnetic order at a finite N\'eel temperature $T_N \sim 200 \, \mathrm{K}$~\cite{PhysRevLett.60.1057}, below which the correlation length $\xi$ diverges in the thermodynamic limit.
Therefore, $\xi$ in our 2D model is much smaller than that in La$_2$CuO$_4$ at the corresponding temperatures, and constrains the mean-free path for magnons in our results.
Since $T_N\sim 200\, \mathrm{K}$ for La$_2$CuO$_4$ is close to the temperature of the anomalous peak in $\kappa$ for La$_2$CuO$_4$ $\sim 300 \, \mathrm{K}$ and marginally close to the lowest temperatures in this paper, an order of magnitude difference in the mean-free path or $\kappa$ appears reasonable.
Nevertheless,  our results demonstrate an increasing $\kappa$ with decreasing $T$, consistent with the high-temperature side of the anomalous peak in $\kappa$ for insulating cuprates.
This feature sits at a similar temperature scale with the anomalous peak in La$_2$CuO$_4$, providing direct evidence that the peak should be attributed to magnons.
For our model in the thermodynamic limit, without constraints of a finite lattice size, $\kappa$ would diverge as $T \rightarrow 0$ with an increasing correlation length and decreasing magnon-magnon scattering.
However, in La$_2$CuO$_4$, a peak appears as temperature decreases and the correlation length is cut off by the onset of long-range order, saturating to the order of the physical system size below the N\'eel temperature. In addition, the mean-free path is affected by other scattering effects, such as phonons and disorder, but these are beyond the scope of our model.

In summary, we provide numerically exact, unbiased results for the specific heat and  thermal conductivity for the half-filled Hubbard model.
From our results based on the Kubo formula without use of Boltzmann theory or any other simplified assumptions, we observe heat conductance by magnons in the Mott insulating antiferromagnetic phase. 
We observe an upturn in $\kappa$, consistent with the high-temperature side of the anomalous peak in thermal conductivity in undoped cuprates.
From the analysis of thermal diffusivity and its temperature and lattice size dependence, we identify different scattering mechanisms that affect magnon heat conductance.

In this paper, we focus on half-filling and the longitudinal magnon heat conductance.
Inspired by our observations, an important open question is: How do magnons impact transport as antiferromagnetism is weakened by factors such as doping and magnetic field?
Further investigation into these effects for both the longitudinal and transverse thermal (Hall) conductivity and the relationship between thermal and charge diffusivity would shed additional light on the nature of heat transport in strongly correlated systems.

\section{Data availability}

The data and analysis routines (Jupyter/Python) needed to reproduce the figures can be found at  \url{https://doi.org/10.5281/zenodo.5305985}.

\begin{acknowledgments}
We acknowledge helpful discussions with R.~T.~Scalettar, S.~Kivelson, A.~Kampf and Y.~Schattner.
\emph{Funding:}
This work was supported by the U.S. Department of Energy (DOE), Office of Basic Energy Sciences,
Division of Materials Sciences and Engineering. 
E.W.H. was supported by the Gordon and Betty Moore Foundation EPiQS Initiative through the grants GBMF 4305 and GBMF 8691.
Computational work was performed on the Sherlock cluster at Stanford University and on resources of the National Energy Research Scientific Computing Center, supported by the U.S. DOE, Office of Science, under Contract no. DE-AC02-05CH11231.
\end{acknowledgments}

\bibliography{main}
\end{document}

\title{Supplementary Material for ``Magnon heat transport in a 2D Mott insulator''}

\author{Wen O. Wang}
\email{wenwang.physics@gmail.com}
\affiliation{Department of Applied Physics, Stanford University, Stanford, CA 94305, USA}

\author{Jixun K. Ding}
\affiliation{Department of Applied Physics, Stanford University, Stanford, CA 94305, USA}

\author{Brian Moritz}
\affiliation{Stanford Institute for Materials and Energy Sciences,
SLAC National Accelerator Laboratory, 2575 Sand Hill Road, Menlo Park, CA 94025, USA}

\author{Edwin W. Huang}
\affiliation{Department of Physics and Institute of Condensed Matter Theory, University of Illinois at Urbana-Champaign, Urbana, IL 61801, USA}

\author{Thomas P. Devereaux}
\email{tpd@stanford.edu}
\affiliation{Stanford Institute for Materials and Energy Sciences,
SLAC National Accelerator Laboratory, 2575 Sand Hill Road, Menlo Park, CA 94025, USA}
\affiliation{
Department of Materials Science and Engineering, Stanford University, Stanford, CA 94305, USA}
\date{\today}

\maketitle

\section{Formalism} \label{formula}
In this section, we provide derivations for the specific heat $c_a$, the thermal conductivity $\kappa_a$, and the electric conductivity $\sigma$.
Here, $\hbar$ and $k_B$ are set to $1$ for convenience, but reintroduced when showing results for $\sigma$, $\kappa_a$ and $c_a$ in the figures, both within the main text and supplementary material.

We consider specific heat in the grand canonical ensemble,
\begin{align}
    c_a = \frac{\dd}{\dd T} \frac{\ev{H_a}}{V}  = -\beta^2 \frac{\dd}{\dd \beta} \frac{\tr H_a e^{-\beta (H-\mu N)}}{V\tr e^{-\beta (H-\mu N)}}, 
\end{align}
where $H$ is the Hamiltonian, $\beta=1/T$ is the inverse temperature, $\mu$ is the chemical potential, $N$ is the total number of particles in the system, and $V$ is the volume of the system. Here, we separate the Hamiltonian into two parts, where $a$ can be taken as $K$ or $P$, representing the kinetic and potential energy terms of the Hamiltonian, or combined for the full Hamiltonian and total energy. 
The kinetic and potential parts of the specific heat, $c_K$ and $c_P$, are defined as the derivative of the corresponding energy with respect to temperature.
We have
\begin{equation}
\frac{\dd}{\dd \beta} e^{-\beta (H-\mu N)} = (-H + N \frac{\dd (\beta \mu)}{\dd \beta})e^{-\beta (H-\mu N)}.
\end{equation}
So
\begin{align}
    &\frac{\dd}{\dd \beta} \ev{H_a} \nonumber \\
    &= \ev{H_a(-H+N\frac{\dd(\beta \mu)}{\dd \beta})} - \ev{H_a}\ev{-H+N\frac{\dd(\beta\mu)}{\beta}} \nonumber \\
    &= -(\ev{H_a H}-\ev{H_a}\ev{H}) + \frac{\dd(\beta\mu)}{\dd\beta} (\ev{H_a N}-\ev{H_a}\ev{N}). \label{component1}
\end{align}
$\mu$ is tuned so that the total density remains fixed while varying the temperature. So
\begin{align}
    & 0=\frac{\dd}{\dd \beta} \ev{N} \nonumber \\
    &= -(\ev{NH}-\ev{N}\ev{H})+\frac{\dd (\beta\mu)}{\dd \beta}(\ev{N^2} - \ev{N}^2).  \label{component2}
\end{align}
Defining $\chi_{O_1 O_2} = \beta(\ev{O_1 O_2}-\ev{O_1}\ev{O_2})$ in Eqs.~\ref{component1} and \ref{component2}, we obtain
\begin{equation}
    c_a = \frac{\beta}{V} (\chi_{H_a H} - \frac{\chi_{H_a N}\chi_{HN}}{\chi_{NN}}).
    \label{specificheatformula}
\end{equation}
Thus, we have two methods to calculate specific heat. One is to measure energies at different temperatures and directly calculate $\delta (E/V)/\delta T$ by choosing a reasonable finite $\delta T$. In other words, we choose two adjacent temperatures $T_1$ and $T_2$ and measure the energy difference, $c_a((T_1+T_2)/2) = \delta \ev{H_a}/\delta T/V = (\ev{H}(T_1)-\ev{H}(T_2))/(T_1-T_2)/V$. The other method is through Eq.~\ref{specificheatformula}, which we term the fluctuation method. In theory they should be identical as $\delta T \rightarrow 0$.
However, in the first method, based on finite differences, if $\delta T$ is too small, $c_a$ will be affected significantly by statistical errors in the energies.
On the other hand, if the finite $\delta T$ is too large, it introduces an additional source of systematic error.  The second method, based on fluctuations, suffers from Trotter error at low temperatures, which will introduce a divergence in $c_a$ as $T\rightarrow 0$~\cite{PhysRevB.36.3833}. (See section~\ref{Trotter})

For heat transport, we consider the response due to a temperature gradient $\nabla_{\gamma} T$ and electric field $\mathbf{E}$.
We define $\overline{\mu} = \mu+e^*V$ so that $\nabla \overline{\mu} = \nabla \mu - e^* \mathbf{E}$, where $e^*$ is the particle charge ($e^* = -e$ for electrons).
We define $\mathbf{J}$ to be the particle current operator, and also define $\mathbf{J}_a$, where $a=Q$ represents the heat current operator,  $a=K$ represents the kinetic energy current operator, and $a=P$ represents the potential energy current operator.
The current responses along the $\alpha$ directions are defined as~\cite{Shastry_2008}
\begin{align}
& \langle J_\alpha \rangle/V = -\beta L^{\alpha\gamma}_{11} \partial_{\gamma} \overline{\mu} + L^{\alpha\gamma}_{12}\partial_{\gamma} \beta \nonumber \\
&=-\beta L^{\alpha\gamma}_{11} \partial_{\gamma} \overline{\mu} - L^{\alpha\gamma}_{12}\beta^2 \partial_{\gamma} T,
\label{eq:current} \\
&\langle J_{a,\alpha} \rangle/V =- \beta L^{\alpha\gamma}_{a1} \partial_{\gamma} \overline{\mu} + L^{\alpha\gamma}_{a2}\partial_{\gamma} \beta \nonumber \\
&=-\beta L^{\alpha\gamma}_{a1} \partial_{\gamma} \overline{\mu} - L^{\alpha\gamma}_{a2}\beta^2 \partial_{\gamma} T. \label{eq:heatcurrent} 
\end{align}
Thus, we have the longitudinal thermal and kinetic/potential-energy conductivities $\kappa_a$ in zero magnetic field and zero electrical current,
\begin{align}
    \kappa_a = \frac{\ev{J_{a,\alpha}}/V}{-\partial_\alpha T} = \beta^2 (L_{a2} - \frac{L_{a1}L_{12}}{L_{11}}), \label{kappa}
\end{align}
and the electric conductivity,
\begin{align}
    \sigma= \frac{e^* \ev{J_{\alpha}}/V}{E_\alpha}
    =-\frac{e^2 \ev{J_{\alpha}}/V}{\partial_{\alpha} \overline{\mu} } = e^2 \beta L_{11}. \label{sigma}
\end{align}
Here the coefficients $L_{a1},L_{12},L_{11}$ are all longitudinal, which means $\alpha = \gamma$ in Eqs~\ref{eq:current} and \ref{eq:heatcurrent}.
The linear response coefficients can be obtained by using perturbation theory~\cite{mahan,Shastry_2008,PhysRev.135.A1505}, 
\begin{align}
    L_{11}(\omega) &= \frac{1}{V \beta} \int_0^\infty \dd \tilde{t} e^{i\tilde{z} \tilde{t}}\int_0^\beta\dd\tau\ev{J_x(\tilde{t}-i\tau)J_x(0)}, \label{definition1} \\
    L_{a1}(\omega) &= \frac{1}{ V \beta}\int_0^\infty \dd \tilde{t} e^{i\tilde{z} \tilde{t}}\int_0^\beta\dd\tau\ev{J_{a,x}(\tilde{t}-i\tau)J_x(0)}, \label{definition2} \\
     L_{12}(\omega) &=\frac{1}{V \beta} \int_0^\infty \dd \tilde{t} e^{i\tilde{z} \tilde{t}}\int_0^\beta\dd\tau\ev{J_{x}(\tilde{t}-i\tau)J_{Q,x}(0)},  \label{definition3} \\
     L_{a2}(\omega) &= \frac{1}{V \beta}\int_0^\infty \dd \tilde{t} e^{i\tilde{z} \tilde{t}}\int_0^\beta\dd\tau\ev{J_{a,x}(\tilde{t}-i\tau)J_{Q,x}(0)},  \label{definition4}
\end{align}
where $\tilde{z}=\omega+i0^+$ and $\tilde{t}$ is the real time.
For operator $O$, 
\begin{equation}
O(\tilde{t}-i\tau) = e^{i(H-\mu N)(\tilde{t}-i\tau)} O e^{-i(H-\mu N)(\tilde{t}-i\tau)} 
\end{equation}
Eqs.~\ref{definition1}-\ref{definition4} are Kubo formulas for the coefficients.
We can find that in Eqs.~\ref{definition1}-\ref{definition4}, replacing $J_{K,x}$ by $J_{K,x}+\lambda_K J_{x}$ and replacing $J_{P,x}$ by $J_{P,x}+\lambda_P J_{x}$ for arbitrary constants $\lambda_K$ and $\lambda_P$, leaves the results unchanged for $\kappa_a$ in Eq.~\ref{kappa} and $\sigma$ in Eq.~\ref{sigma}.
In our calculations, we measure the correlators in imaginary time $\ev{T_\tau O_1(\tau)O_2(0)}\equiv Z^{-1}\mathrm{Tr}(e^{-\beta(H-\mu N)} T_\tau e^{\tau (H-\mu N)}O_1 e^{-\tau (H-\mu N) } O_2)$, where $Z=\mathrm{Tr}(e^{-\beta(H-\mu N)})$ is the partition function, and obtain the coefficients in real frequency using maximum entropy (MaxEnt) analytic continuation~\cite{ana1,ana2}, where details of the method are described in Ref~\cite{Huang987}.

We consider the Hubbard model,
\begin{align}
H = - &t \sum_{\ev{lp}\sigma} \l(c^\dagger_{l\sigma}c_{p\sigma}
 + c^\dagger_{p\sigma}c_{l\sigma}\r) 
  \nonumber \\ &
 + U\sum_{l} \l(n_{l\uparrow}-\frac{1}{2}\r)\l(n_{l\downarrow}-\frac{1}{2}\r),
\label{hubbard}
\end{align}
where $t$ is the nearest-neighbour hopping, $U$ is the on-site Coulomb interaction,
$\mathit{c}_{l,\mathit{\sigma}}^{\dagger}$ $(\mathit{c}_{l,\mathit{\sigma}})$ is the creation (annihilation) operator for an electron at site $l$ with spin $\mathit{\sigma}$, and $\mathit{n}_{l,\mathit{\sigma}} \equiv \mathit{c}_{l,\mathit{\sigma}}^{\dagger} \mathit{c}_{l,\mathit{\sigma}}$ is the number operator at site $l$. 
The model is placed on a square lattice with periodic boundary conditions.
It can be separated easily into kinetic $\propto t$ and potential $\propto U$ contributions.

Therefore the local energy at site $l$ is
\begin{align}
& h_l = - \frac{t}{2} \sum_{\bm{\delta},\sigma} \l(c^\dagger_{l+\bm{\delta},\sigma}c_{l,\sigma}
 + c^\dagger_{l,\sigma}c_{l+\bm{\delta}\sigma}\r) \nonumber \\ &
 + U \l(n_{l\uparrow}-\frac{1}{2}\r)\l(n_{l\downarrow}-\frac{1}{2}\r),
\end{align}
where $\bm{\delta}$ includes all position displacements for nearest neighbours. Specifically, on a 2D square lattice $\bm{\delta}=+\mathbf{x}, -\mathbf{x}, +\mathbf{y}, -\mathbf{y}$, where the lattice constant is set to $1$ and $\mathbf{x}$ and $\mathbf{y}$ are unit vectors. 

We define the chemical potential as $\mu$, the position of site $l$ as $\mathbf{r}_l$ and $\mathbf{R}_E = \sum_l \mathbf{r}_l h_l$.
Since the total energy is conserved, the energy current $\mathbf{J}_E$ using the equation of continuity is~\cite{mahan}
\begin{align}
&\mathbf{J}_E  = i[H-\mu N, \mathbf{R}_E] 
= i\sum_{p,l} [h_p-\mu n_p, \mathbf{r}_l h_l] \nonumber\\
&= \sum_{l,\bm{\delta}_1,\bm{\delta}_2 \in \{\bm{\delta}\},\sigma}(-\frac{\bm{\delta}_1+\bm{\delta}_2}{4}) {t^2}(i c^{\dagger}_{l+\bm{\delta}_1+\bm{\delta}_2,\sigma}c_{l,\sigma} + h.c.) \nonumber\\
&+\frac{Ut}{4} \sum_{l,\bm{\delta} \in \{\bm{\delta}\},\sigma} \bm{\delta} (n_{l+\bm{\delta},\bar{\sigma}}+n_{l,\bar{\sigma}})(ic_{l+\bm{\delta},\sigma}^\dagger c_{l,\sigma}+h.c.) \nonumber\\
&- \frac{Ut}{4} \sum_{l,\bm{\bm{\delta}} \in \{\bm{\delta}\},\sigma}\bm{\delta}
(ic^\dagger_{l+\bm{\delta},\sigma}c_{l,\sigma}+h.c.). 
\label{energycurrent}
\end{align}
Here the spin index $\bar{\sigma}=-\sigma$.
In order to separate the kinetic part and potential part in $\kappa$ we need to define the kinetic energy current operator $\mathbf{J}_K$ and potential energy current operator $\mathbf{J}_P$.
However, we cannot use the equation of continuity $\partial H_{K/P}/\partial t = -\nabla \cdot \mathbf{J}_{K/P}$ to define them since  $H_{K/P}$ are not themselves conserved quantities. 
Instead, $\mathbf{J}_E$ in Eq.~\ref{energycurrent} can be readily separated into two parts. 
We defined the two-fermion term $\propto t^2$ as the kinetic energy current $\mathbf{J}_K$
and the four-fermion term $\propto Ut$  as the potential energy current $\mathbf{J}_P$, 
in analogy to how the Hamiltonian is split into $H_K$ and $H_P$.
Therefore we can define the longitudinal kinetic/potential-energy conductivities $\kappa_{K/P} = -\ev{J_{K/P,x}}/\partial_x T$ along the $x$ direction. 
As mentioned in the main text, from Eq.~\ref{energycurrent}, operators in $\mathbf{J}_{K}$ involve next-nearest and next-next-nearest-neighbor hopping, which for an antiferromagnetic spin pattern are forbidden by Pauli exclusion, as spins at the two sites sit on the same sublattice and therefore have the same orientation.
On the other hand, terms in the $\mathbf{J}_{P}$ are allowed at the expense of forming double occupancies, which cost an energy $\sim U$.

The heat current is determined by
\begin{equation}
    \mathbf{J}_Q = \mathbf{J}_E - \mu \mathbf{J},
\end{equation}
where $\mathbf{J}$ is the particle current,
\begin{align*}
 &  \mathbf{J} = i[H-\mu N, \mathbf{R}_N] = i\sum_{p,l} [h_p-\mu n_p, \mathbf{r}_l n_l] \\
&=\frac{t}{2}\sum_{l,\bm{\delta} \in \{\bm{\delta}\},\sigma}\bm{\delta}
(ic^\dagger_{l+\bm{\delta},\sigma}c_{l,\sigma}+h.c.).
\end{align*}

At half-filling, $\mu=0$.
We consider the case of a bipartite cluster divided into two sublattices, A and B, where no hopping occurs between sites on the same sublattice for the Hamiltonian in Eq.~\ref{hubbard}.
The particle-hole transformation can be defined as
\begin{eqnarray}
d_{l}=\tilde{\alpha}_l c_{l,\sigma}^\dagger,
\end{eqnarray}
where $\tilde{\alpha}_l = 1$ for $l\in A$ and $\tilde{\alpha}_l = -1$ for $l\in B$.
We thus have particle-hole symmetry in the grand canonical Hamiltonian $H-\mu N$ at half-filling, which means, after the transformation, the Hamiltonian remains unchanged with that appearing in Eq.~\ref{hubbard}, only with $c \rightarrow d$.
It is notable that in Eq.~\ref{energycurrent}, $c_{l+\bm{\delta}_1+\bm{\delta}_2}$ and $c_{l}$ are always on the same sublattice, while $c_{l+\bm{\delta}}$ and $c_{l}$ are always on different sublattices.
So after the particle-hole transformation, $\mathbf{J}_K \rightarrow -\mathbf{J}_K$, $\mathbf{J}_P \rightarrow -\mathbf{J}_P$,  $\mathbf{J} \rightarrow \mathbf{J}$, with the operator substitution $c \rightarrow d$.
So we know that particle-hole symmetry guarantees $\ev{T_\tau \mathbf{J}_K(\tau)\mathbf{J}} = \ev{T_\tau \mathbf{J}_P(\tau)\mathbf{J}} =0$.
This leads to $L_{a1}=L_{12} = 0$ in Eqs.~\ref{definition2} and ~\ref{definition3}.
To measure $\kappa_a$ and $\sigma$, we only need to measure and apply analytic continuation to imaginary-time correlators $\ev{T_\tau J_{x}(\tau)J_{x}(0)}$, $\ev{T_\tau J_{K,x}(\tau)J_{K,x}(0)}$, $\ev{T_\tau J_{P,x}(\tau)J_{P,x}(0)}$, and $\ev{T_\tau J_{E,x}(\tau)J_{E,x}(0)}$.  The real parts of the coefficients $\Re L_{11}(\omega)$ or $\Re L_{aa}(\omega)$ are guaranteed to be positive definite (see Eq.~\ref{whyeven} in section~\ref{sec:parameter}). 
Other coefficients needed in $\kappa_a$ can be obtained by linear combinations of these coefficients.
Similarly, one can verify that $\chi_{H_aN}=0$. So in Eq.~\ref{specificheatformula}, we only need to calculate the energy fluctuation function $\chi_{H_aH}$ for the specific heat.

Here, we discuss a limitation of our methods for analytic continuation.
The kernel, which converts the imaginary-time correlator into real frequency, limits the frequency resolution 
to $\sim T$~\cite{Huang987,PhysRevResearch.3.033033}.
If the scattering rate decays faster than $T$ as temperature decreases, 
for example $\sim T^2$ as suggested by other work~\cite{PhysRevB.42.2096},
the Drude peak of $\Re \kappa(\omega)$ may become resolution limited,
making it difficult to 
capture the proper frequency dependence around $\omega=0$ at low temperatures; although future work would be necessary to verify this scenario.

\begin{figure*}
    \centering
    \includegraphics[width=\textwidth]{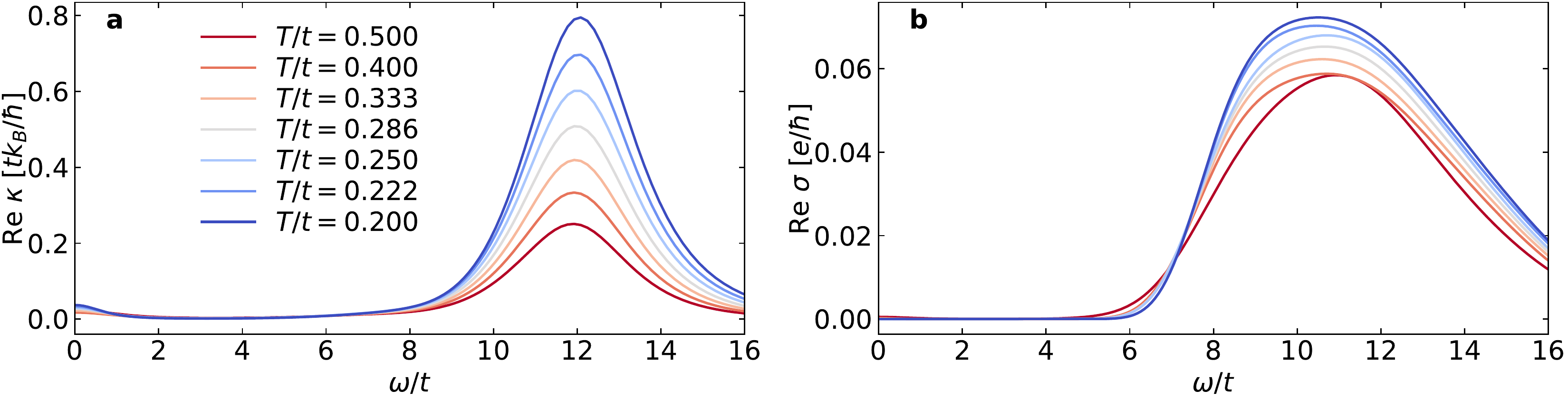}
    \caption{Profiles of $\Re \kappa(\omega)$(\textbf{a}) and  $\Re \sigma(\omega)$(\textbf{b}) over a wide frequency range. Parameters and methods are the same as Fig.~1 in the main text.
    }
    \label{fig:frequency_whole}
\end{figure*}

\section{Trotter error and finite-size effect} \label{Trotter}

\begin{figure}
    \centering
    \includegraphics[width=\textwidth]{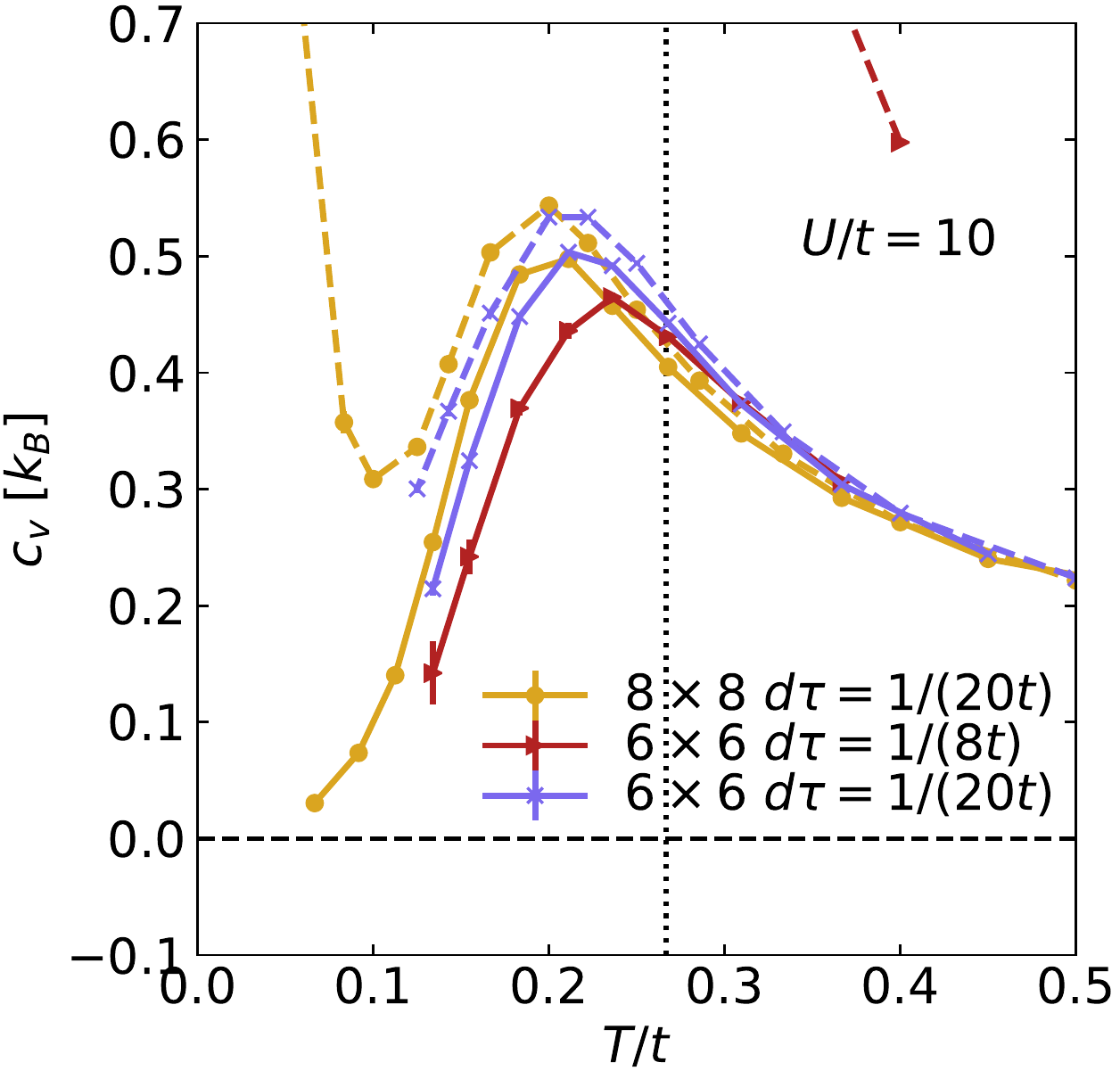}
    \caption{Specific heat $c_v$ calculated with three different sets of parameters including lattice sizes and $\dd\tau$. Solid lines: calculated by the method of finite differences. 
    Dashed lines: calculated from energy fluctuation as in Eq.~\ref{specificheatformula}.
    Error bars (mostly smaller than the marker size) represent $\pm 1$ standard error  determined by jackknife resampling~\cite{jackknife}.
    The black dotted line indicates the temperature $T=2J/3$.
    }
    \label{fig:finite-size}
\end{figure}

For completeness, we show profiles of $\Re \kappa(\omega)$ and $\Re \sigma(\omega)$ over a wide frequency range in Fig.~\ref{fig:frequency_whole}, with dominant high-frequency peaks at $T\sim U$. For the specific heat, as shown in previous work using QMC~\cite{PhysRevB.63.125116,PhysRevB.55.12918}, low-temperature peak positions for the Hubbard model with $U/t\geq 10$ match closely to $2J/3$, the peak position numerically calculated for the Heisenberg model~\cite{PhysRevLett.77.892}.
However, our result in Fig.~3 in the main text deviates from this position even for $U/t=10$ and $12$, as we simulate on a larger lattice size and for a smaller imaginary time discretization $\dd\tau$.
The difference between our result and that in previous work is a combined effect of finite-size and the Trotter error, as shown in Fig.~\ref{fig:finite-size}.
The red solid line shows results using the same lattice size and $\dd \tau$ as Ref.~\cite{PhysRevB.63.125116}. 
Changes in lattice size and $\dd\tau$ both change the peak position to some extent; however, the qualitative behavior does not change.
The deviation from $2J/3$ also stems from the effective spin exchange energy $J^*$ for Hubbard model with intermediate $U$, which can deviate from the leading order approximation $J=4t^2/U$~\cite{Huang1161, Huang2018}.
As a point of comparison, this observed deviation in the peak position agrees roughly with Ref.~\cite{PhysRevA.86.023633} using the numerical linked-cluster expansion (NLCE)~\cite{PhysRevLett.97.187202,TANG2013557,PhysRevA.84.053611,Nichols383}, which is free of finite-size effects and Trotter error. 

\begin{figure*}
    \centering
    \includegraphics[width=\textwidth]{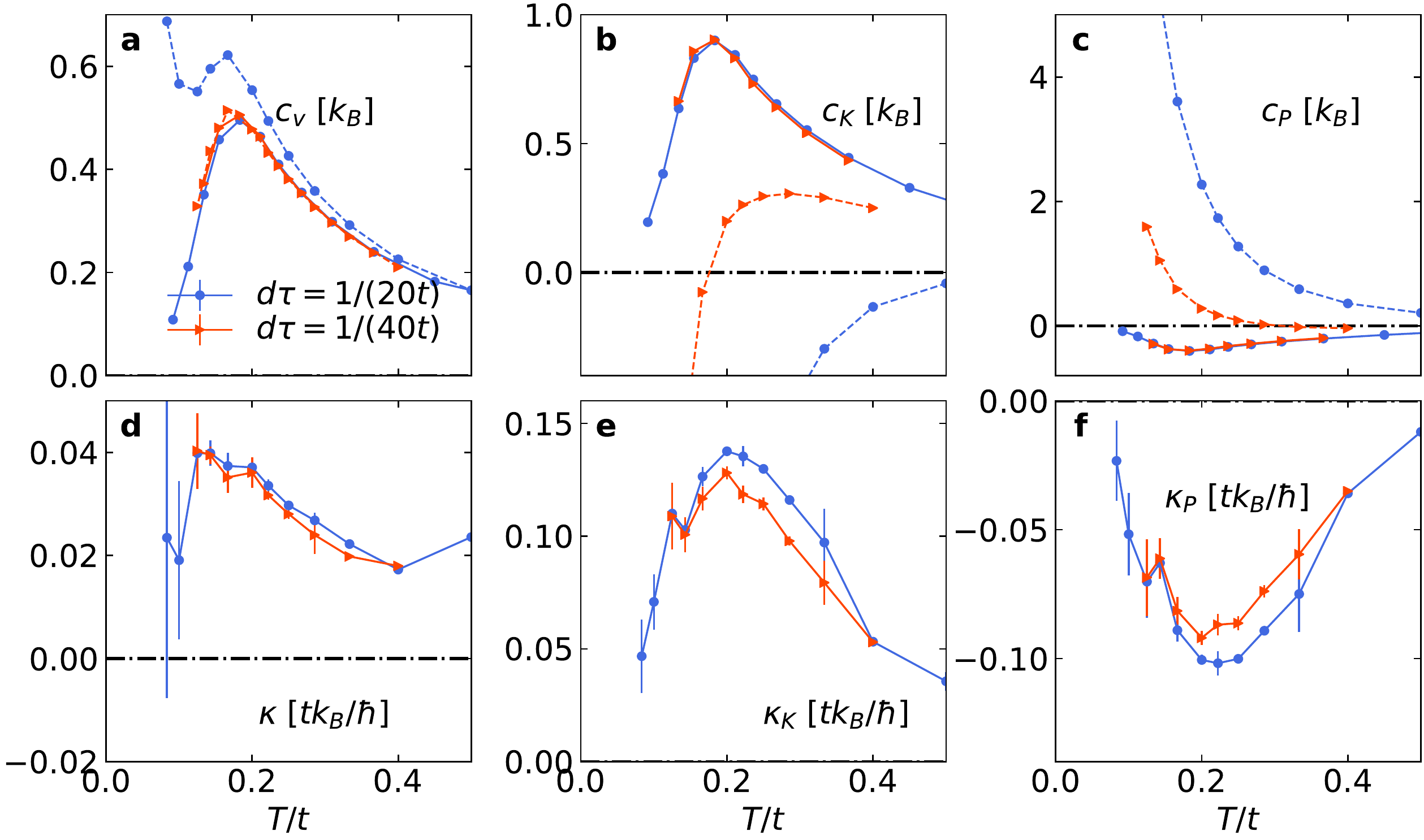}
    \caption{Trotter error analysis. (\textbf{a}) $c_v$,  (\textbf{b}) $c_K$, (\textbf{c}) $c_P$, (\textbf{d}) $\kappa$, (\textbf{e}) $\kappa_K$, and (\textbf{f}) $\kappa_P$ calculated with imaginary time discretization $\dd\tau=1/(20t)$ and $1/(40t)$ on lattices of size $8\times8$ for $U/t=12$.
    In panels \textbf{a}-\textbf{c}, solid lines are calculated by the method of finite differences, while dashed lines are calculated from energy fluctuation as in Eq.~\ref{specificheatformula}.  For panels \textbf{a}-\textbf{c}, error bars (smaller than the marker size) represent $\pm 1$ standard error determined by jackknife resampling.
    For panels \textbf{d}-\textbf{f}, error bars represent $1$ bootstrap standard error~\cite{bootstrap}.
    }
    \label{fig:trotter}
\end{figure*}

Looking more closely at the Trotter error, the conventional constraint for a sufficiently small $\dd\tau$ requires $U(\dd\tau)^2 \leq 1/(8t) = 1/W$, where $W=8t$ is the non-interacting bandwidth. However, this constraint is far from sufficient for the fluctuation method based on Eq.~\ref{specificheatformula}, as shown in Fig.~\ref{fig:trotter}.
For the largest value of the interaction in our study $U/t=12$ we find that even $\dd\tau = 1/(20t)$ is not small enough to sufficiently suppress Trotter error at low temperatures $T/t<0.2$, as shown in Fig.~\ref{fig:trotter} (\textbf{a}).
Further, the highest temperature where $c_K$ and $c_P$ calculated from the fluctuation method with $\dd\tau = 1/(20t)$ start to show significant trotter error is even higher than that for $c_v$, as shown in Fig.~\ref{fig:trotter} (\textbf{a})-(\textbf{c}).
At the lowest temperatures for $\dd\tau = 1/(20t)$,  $c_v$, $c_K$ and $c_P$ calculated in this way diverge as $T\rightarrow 0$~\cite{PhysRevB.36.3833}.
However, $c_v$, $c_K$ and $c_P$ calculated from finite energy differences have far less Trotter error for the same $\dd\tau$. 
As shown in Fig.~\ref{fig:trotter} (\textbf{d})-(\textbf{f}),
the change in  $\dd\tau$ from $1/(20t)$ to $1/(40t)$  does not qualitatively affect the results for $\kappa$, $\kappa_K$, and $\kappa_P$.

\begin{figure*}
    \centering
    \includegraphics[width=\textwidth]{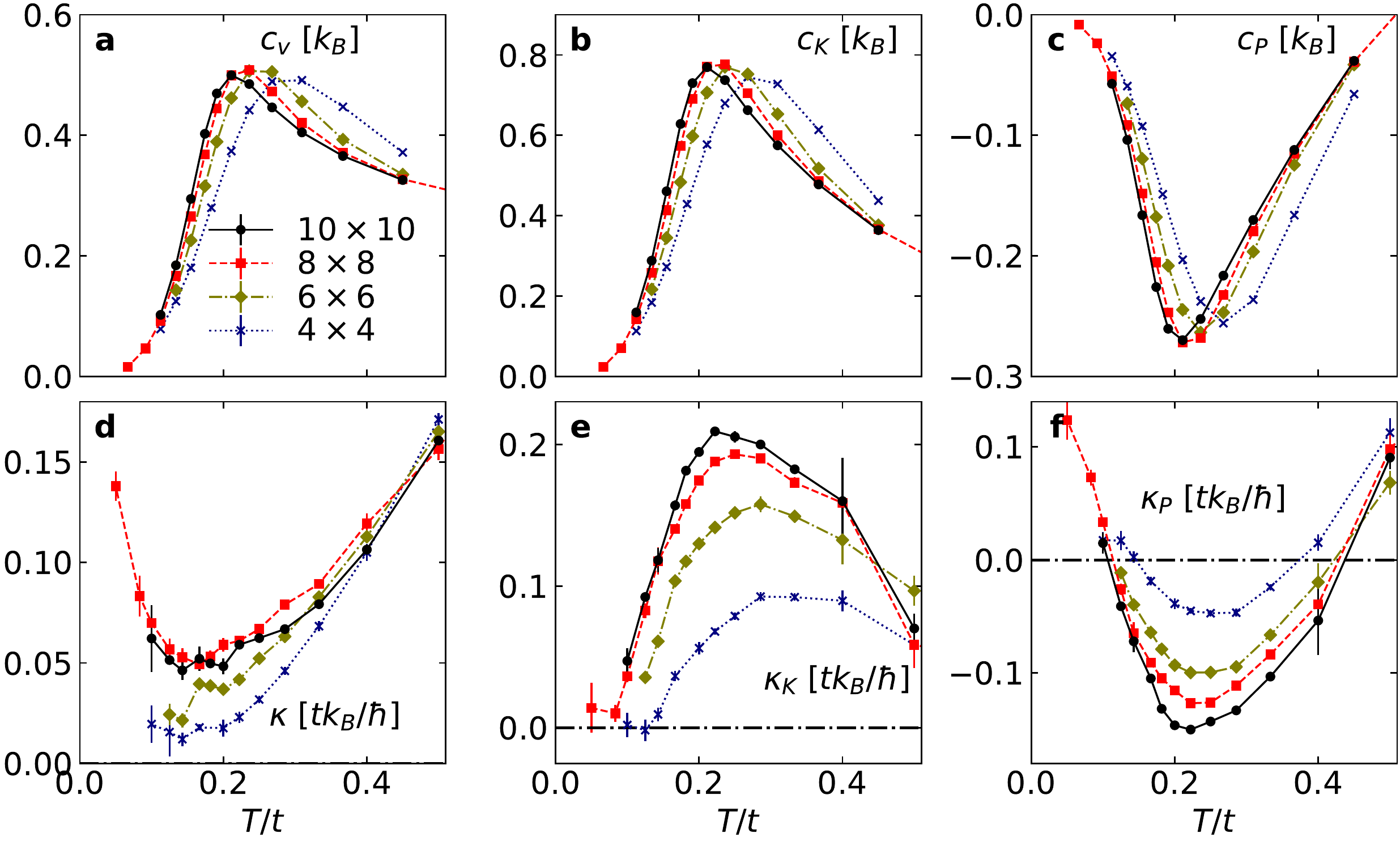}
    \caption{Finite-size analysis. (\textbf{a}) $c_v$ ,  (\textbf{b}) $c_K$, (\textbf{c}) $c_P$, (\textbf{d}) $\kappa$, (\textbf{e}) $\kappa_K$, and (\textbf{f}) $\kappa_P$ calculated for lattices of size $4 \times 4$, $6 \times 6$,  $8 \times 8$ and $10 \times 10$, with $U/t=8$ and an imaginary time discretization $\dd\tau=1/(20t)$. $c_v$, $c_K$ and $c_P$ are calculated by the method of finite differences. For panels \textbf{a}-\textbf{c}, error bars (smaller than the markers) represent $\pm 1$ standard error determined by jackknife resampling.
    For panels \textbf{d}-\textbf{f}, error bars represent $1$ bootstrap standard error.
    }
    \label{fig:finitesize_U8}
\end{figure*}

\begin{figure*}
    \centering
    \includegraphics[width=\textwidth]{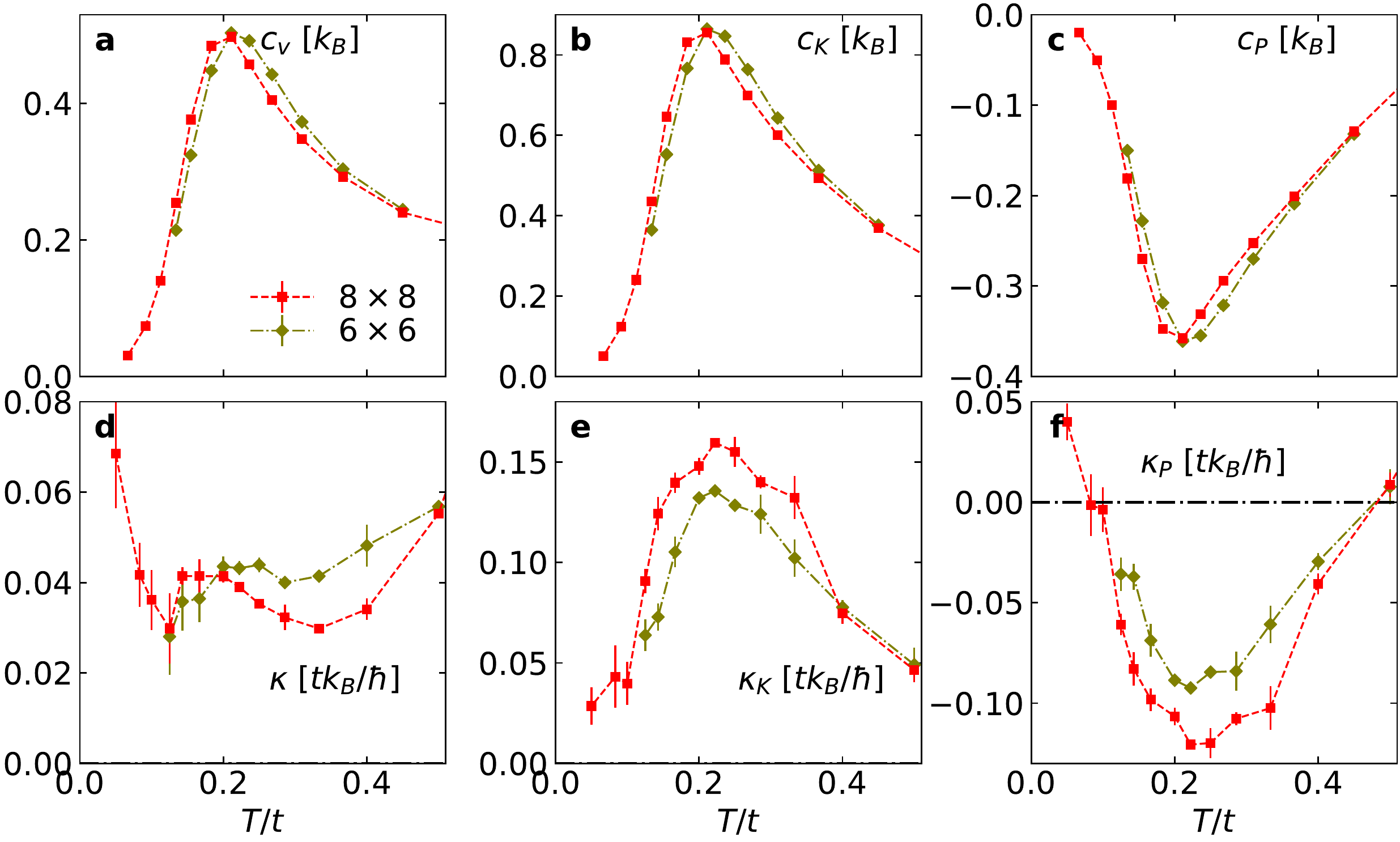}
    \caption{Similar finite-size analysis as in Fig.~\ref{fig:finitesize_U8} but for $U/t=10$. Lattice sizes include $6 \times 6$ and $8 \times 8$.
    }
    \label{fig:finitesize_U10}
\end{figure*}

\begin{figure*}
    \centering
    \includegraphics[width=\textwidth]{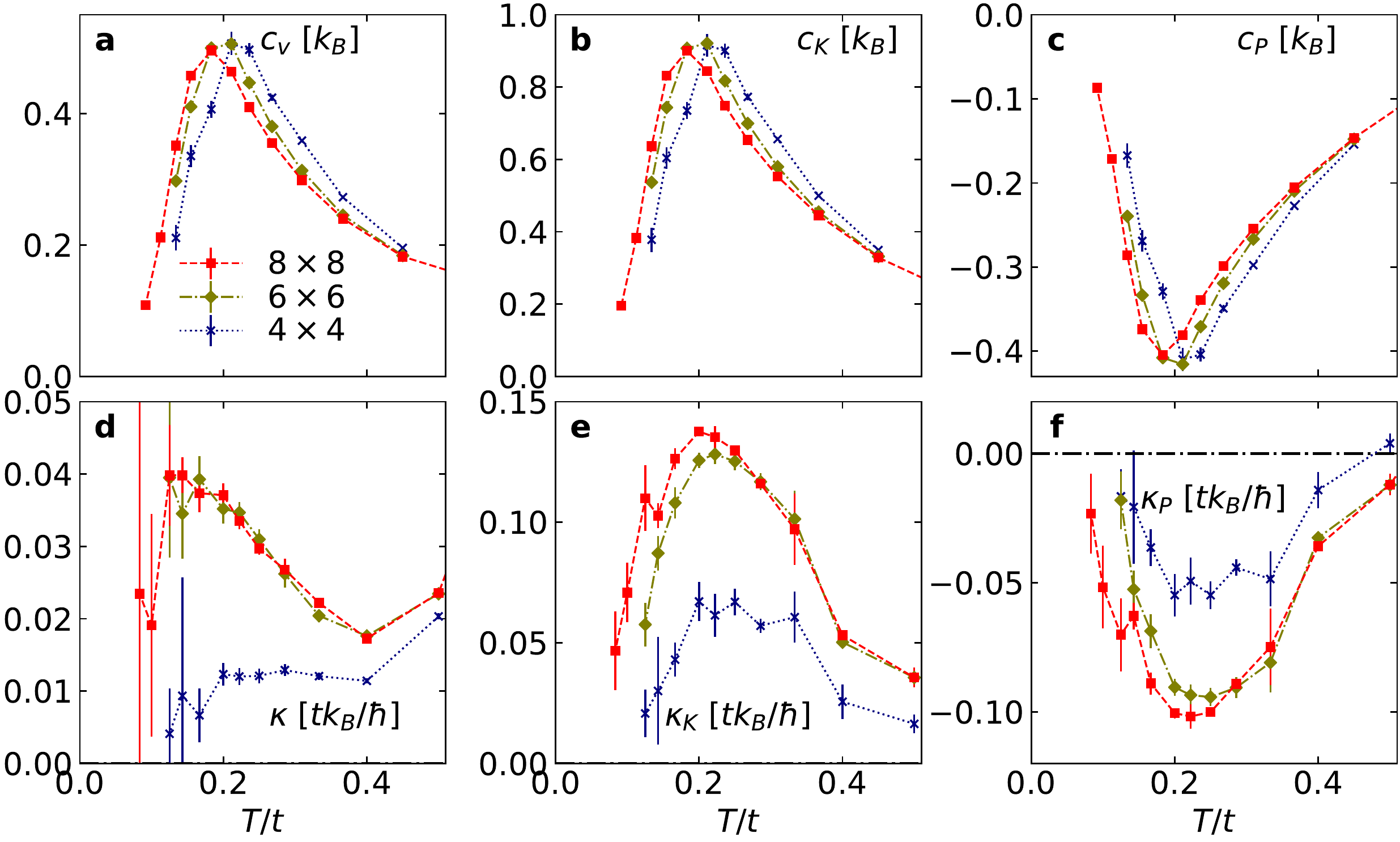}
    \caption{Similar finite-size analysis as in Fig.~\ref{fig:finitesize_U8} but for $U/t=12$. Lattice sizes include $4 \times 4$,  $6 \times 6$, and  $8 \times 8$.
    }
   \label{fig:finitesize_U12}
\end{figure*}

In Fig.~\ref{fig:finitesize_U8}, we show the finite-size analysis for $U/t=8$.
We find that the finite-size does not affect the overall qualitative behavior of either the specific heat or thermal conductivity.
Quantitatively, the temperatures for the magnon peaks in $c_v$, $c_K$, and $c_P$, shift slightly to lower temperatures for larger lattice sizes.
For $\kappa$, $\kappa_K$, and $\kappa_P$ at the lowest temperatures, the overall trend is that magnitudes increase with larger lattices, except between lattice sizes of $8 \times 8$ and  $10 \times 10$, where these differences are minimal. 
Similar size dependences are shown in Figs.~\ref{fig:finitesize_U10} and \ref{fig:finitesize_U12} for $U/t=10$ and $12$, respectively.

\begin{figure*}
    \centering
    \includegraphics[width=0.9\textwidth]{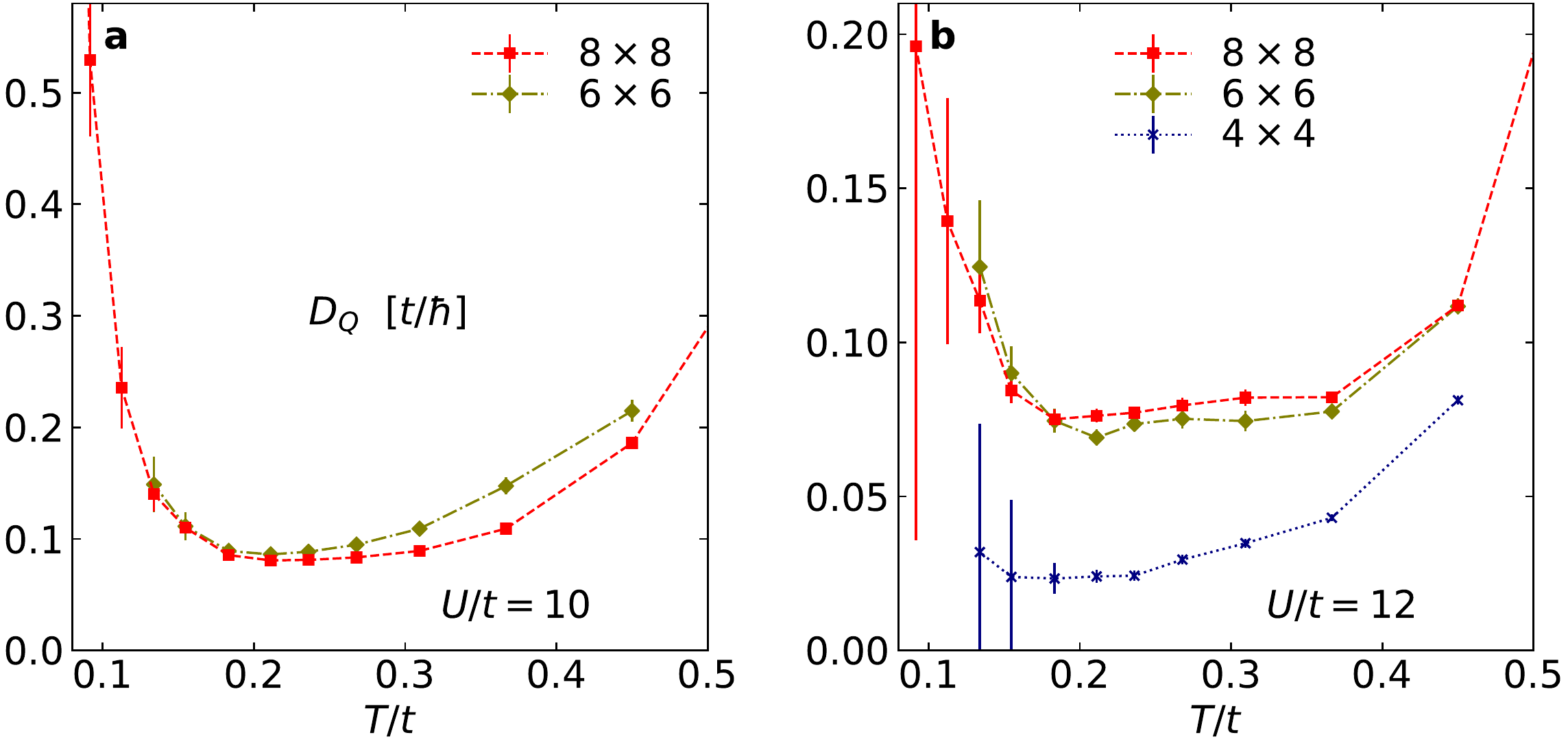}
    \caption{Lattice size dependence of thermal diffusivity $D_Q$ for $U/t=10$ (\textbf{a}) and $U/t=12$ (\textbf{b}).
    Imaginary time discretization $\dd\tau=1/(20t)$.
    Methods are the same as in Fig.~4 in the main text.
    }
   \label{fig:diff_FF}
\end{figure*}

Similar to Fig.~4 (\textbf{b}) in the main text, we show the temperature and lattice size dependence of $D_Q$ for $U/t=10$ and $12$ in Figs.~\ref{fig:diff_FF}(\textbf{a}) and (\textbf{b}), respectively.
Comparing Fig.~\ref{fig:diff_FF} with Fig.~4 (\textbf{b}) in the main text, we notice that for $U/t\geq 10$ at the lowest temperatures, the difference between $D_Q$ on $6\times 6$ and $8\times 8$ lattices is much less obvious than for $U/t=8$. This is because $J$ is larger for smaller $U$. Therefore, the temperature at which the correlation length saturates is larger for smaller $U$~\cite{PhysRevLett.104.066406}.

\begin{figure*}
    \centering
    \includegraphics[width=0.9\textwidth]{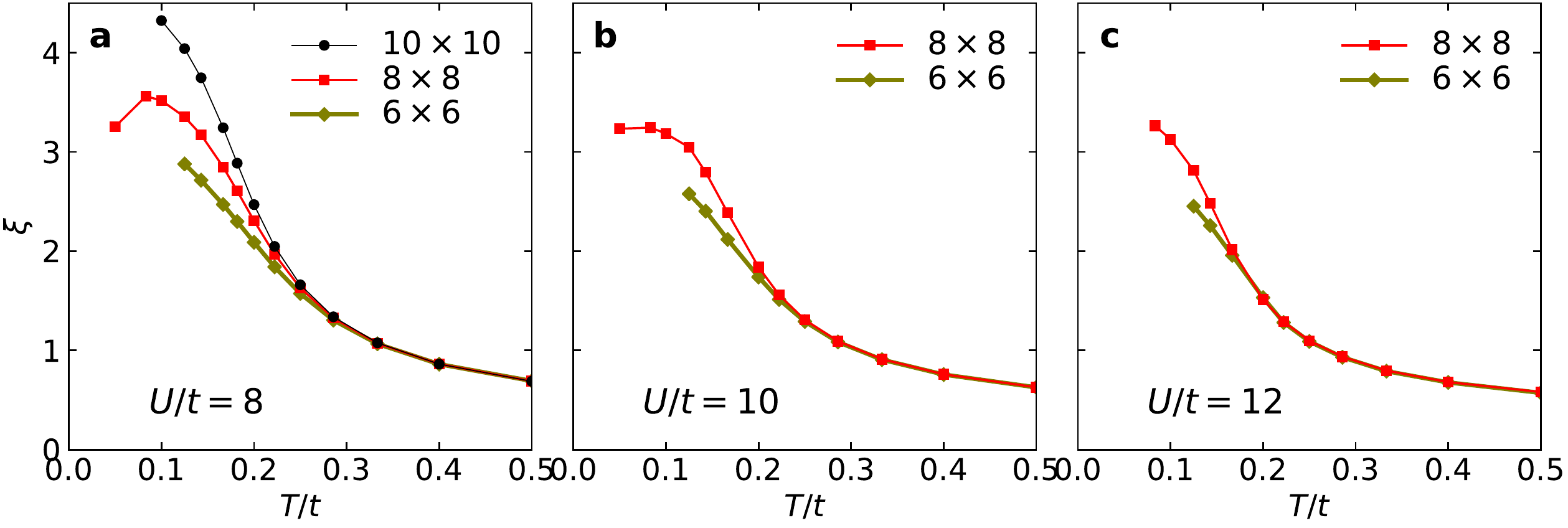}
    \caption{Lattice size and temperature dependence of the spin-spin correlation length $\xi$ for $U/t=8$ (\textbf{a}), $10$ (\textbf{b}), and $12$ (\textbf{c}).
    $\xi$ is approximated by fitting the staggered spin-spin correlation function along the $x$ direction in real space, $(-1)^{\Delta r_x}\ev{S^z(\mathbf{r}+ \Delta r_x \mathbf{x})S^z(\mathbf{r})}$, to $C_{ss}(e^{-|\Delta r_x|/\xi}+e^{-|L_x-\Delta r_x|/\xi})$.
     Here $S^z(\mathbf{r})\equiv (n_{\uparrow}-n_{\downarrow})(\mathbf{r})$ is the $z$-direction component of the spin operator at position $\mathbf{r}$, $\Delta r_x$ is the distance in the $x$ direction, $C_{ss}$ is a fitting parameter, and $L_x$ is the linear size of the lattice along the $x$ direction.
    The correlation function is position independent due to translation symmetry.
    The values at $r_x=0$ and $r_x=1$ are not included in the fitting, in order to rule out short-range effects.
    The term $\propto e^{-|L_x-\Delta r_x|/\xi}$ is added to the fitting function, in consideration of periodic boundary conditions~\cite{PhysRevLett.64.1449}.
    Imaginary time discretization $\dd\tau=1/(20t)$.
    }
   \label{fig:correlationlength}
\end{figure*}

In Fig.~\ref{fig:correlationlength}, we show the lattice size and temperature dependence of the spin-spin correlation length $\xi$. As mentioned in the main text, we find that $\xi$ increases with decreasing temperatures and saturates at some temperature.
This saturation occurs at slightly less than half of the linear size of the lattice, since we use periodic boundary conditions.

\section{Spin-wave theory} \label{sec:spinwavetheory}

In the limit $t/U \ll 1$, for temperature much lower than $U$, the effective model is the spin-$\frac{1}{2}$ antiferromagnetic Heisenberg model~\cite{Auerbach1994heisenberg}.
According to spin-wave theory~\cite{PhysRev.87.568}, which utilizes the Holstein-Primakoff transformation, one can obtain an approximate Hamiltonian, 
\begin{equation}
    H = E_0 + \sum_{\mathbf{k}}\omega_{\mathbf{k}} (\alpha_{\mathbf{k}}^\dagger \alpha_{\mathbf{k}} + \beta_{\mathbf{k}}^\dagger \beta_{\mathbf{k}}),
\end{equation}
where $E_0$ is the ground state energy, and $\alpha_{\mathbf{k}}^\dagger$ and $\beta_{\mathbf{k}}^\dagger$ create magnon excitations with wavevector $\mathbf{k}$ and energy $\omega_{\mathbf{k}}$. 
The dispersion relation for both $\alpha_{\mathbf{k}}$ and $\beta_{\mathbf{k}}$ is
\begin{equation}
    \omega_{\mathbf{k}} = JSz \sqrt{1-\gamma_{\mathbf{k}}^2}, \label{omegaK}
\end{equation}
where the spin exchange energy $J=4t^2/U$, the spin $S=1/2$, and $z$ is the coordination of each site, equal to the number of nearest neighbors on the lattice, or twice the dimension of our square lattice.
$\gamma_{\mathbf{k}}$ is determined by
\begin{equation}
    \gamma_{\mathbf{k}} = \frac{1}{z} \sum_{\bm{\delta}} \cos (\mathbf{k} \cdot \bm{\delta}).
\end{equation}
In deriving this expression, our 2D square lattice has been divided into two interpenetrating sublattices. 
So, $\mathbf{k}$ should be chosen in the magnetic Brillouin zone, which is formed by either sublattice with a lattice constant $\sqrt{2}$, while $\bm{\delta}= +\mathbf{x}, -\mathbf{x}, +\mathbf{y}, -\mathbf{y}$ are the nearest neighbour displacements on the original lattice.
From Eq.~\ref{omegaK} we know that there are two branches of low-energy magnons around $\mathbf{k}=(0,0)$ and $(\pi,\pi)$~\cite{PhysRevB.92.054409}.

\begin{figure}
    \centering
    \includegraphics[width=\textwidth]{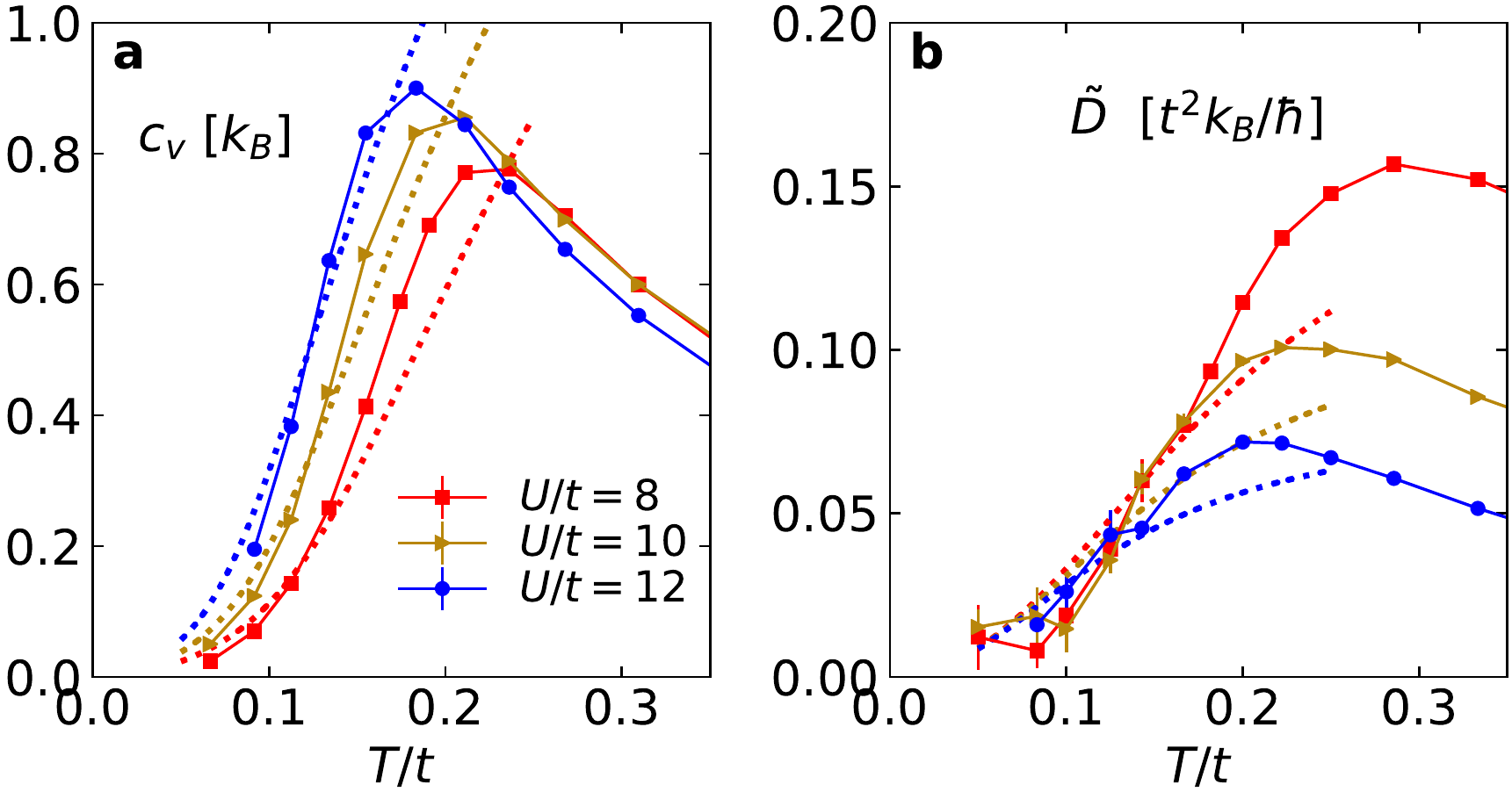}
    \caption{$c_K$ and $\tilde{D}$ compared with predictions from spin-wave theory. (\textbf{a}) Kinetic part of the specific heat $c_K$ (solid lines as in Fig.~3 (\textbf{b})) in the main text. 
    Here the dotted lines show specific heat from the 2D antiferromagnetic Heisenberg model for the corresponding value of $J$, obtained from spin-wave theory. 
    (\textbf{b}) Kinetic part of the thermal Drude weight $\tilde{D}$ in the Hubbard model (solid lines) compared to that obtained from spin-wave theory  and the Drude formula for the 2D antiferromagnetic Heisenberg model (dotted lines). 
    %Methods are in the Supplementary Material~\cite{supplementary}.
    The same color (marker) represents the same $U$.
    The error bars for the Drude weight represent  $1$ bootstrap standard error.
    Simulation lattice size for the Hubbard model is $8 \times 8$, while the calculation for Heisenberg model is in the thermodynamic limit.
    }
    \label{fig:seperatecompare}
\end{figure}

\begin{figure*}
    \centering
    \includegraphics[width=\textwidth]{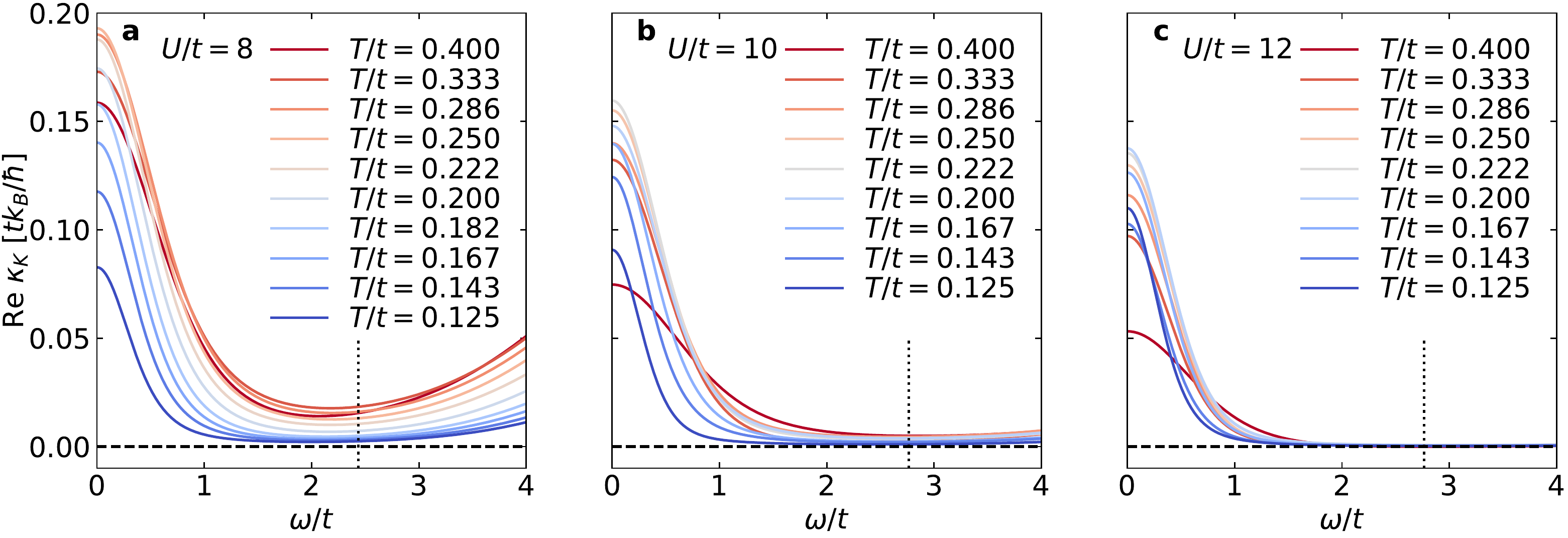}
    \caption{Frequency dependence of $\Re \kappa_K (\omega)$ at different temperatures for different $U$. Vertical dotted lines indicate the upper frequency bound $\Omega$ for calculation of the Drude weight.
    Simulation lattice size is $8 \times 8$.
    Imaginary time discretization $\dd\tau=1/(20t)$.}
    \label{fig:drude}
\end{figure*}

%Now we focus on the quantitative behaviors of $c_v$ and $\kappa$ for $T \sim J$.
We compare our results for Hubbard model with spin-wave theory for non-interacting magnons.
As we discussed in the main text, 
the potential parts $c_P$ and $\kappa_P$ should not be considered, as double occupancies are projected out when the Hubbard model is mapped to the antiferromagnetic Heisenberg model, from which one derives the spin-wave theory.
Therefore, for the temperature regime where the kinetic contribution dominates $\kappa$, a combination of spin-wave theory and the Drude formula provide a reasonable approximation for comparison.
In Fig.~\ref{fig:seperatecompare}, we compare the kinetic parts of the specific heat $c_K$ and the thermal Drude weight $\tilde{D}$ with the predictions of spin-wave theory.
We find that $c_K$ and $\tilde{D}$ agree with the prediction from the spin-wave theory in this temperature regime, consistent with magnons being the dominant excitations and heat carriers. 
The kinetic part of thermal Drude weight~\cite{PhysRevB.66.140406}, defined as $\tilde{D}\equiv\int_0^\Omega \dd \omega \Re \kappa_K(\omega)$, rather than $\kappa_K$ itself, is used for comparison, due to the difficulty in determining the scattering rate of the magnons in the Heisenberg model. Here, $\Omega$ is an appropriate frequency bound to include the whole low-frequency Drude peak.
$\Omega$ should be large enough to include all the low-frequency Drude-like behavior, while it also should be small enough to avoid high frequency behavior, which is dominated by the higher-energy terms integrated-out in deriving the Heisenberg Hamiltonian. Our choices for $\Omega$ and their relative positions in the spectra of $\Re \kappa_K (\omega)$ are shown in Fig.~\ref{fig:drude}.
Below, we will provide details about the methods used to determine $c_v$ and the thermal Drude weight for the Heisenberg model from spin-wave theory.

Dotted lines in Fig.~\ref{fig:seperatecompare}(\textbf{a}) are obtained by calculating the following expression in the thermodynamic limit:
\begin{align}
   &\frac{\partial}{\partial T} E =\frac{1}{V} \sum_{\mathbf{k}}
   2\omega_{\mathbf{k}}\frac{\partial}{\partial T}n(\mathbf{\mathbf{k}}) = \iint \dd\mathbf{k} c_v(\mathbf{k}) \label{cvini} \\
   &=\frac{2}{4\pi^2}\int_0^{\frac{2\pi}{\sqrt{2}}} \dd k_{x'} \int_0^{\frac{2\pi}{\sqrt{2}}}  \dd k_{y'} \frac{2\omega_{\mathbf{k}}^2/T^2 e^{\omega_{\mathbf{k}}/T}}{(e^{\omega_{\mathbf{k}}/T}-1)^2},
\end{align}
where $n(\mathbf{k})$ is the Bose-Einstein distribution $n(\mathbf{k}) = 1/(e^{ \omega_{\mathbf{k}}/ T}-1)$. The factor $2$ in  Eq.~\ref{cvini} comes from the fact that there are two types of magnon excitations $\alpha_{\mathbf{k}}$ and $\beta_{\mathbf{k}}$.
Here, the $x'$ and $y'$ directions are along the next-nearest-neighbour bonds of the original lattice, which are the nearest-neighbours of the sublattice with a distance $\sqrt{2}$.

For dotted lines in Fig.~\ref{fig:seperatecompare}(\textbf{b}), we use the thermal conductivity expression for the bosonic gas derived from the Boltzmann Equation~\cite{dicastro_raimondi_2015, ashcroft1976solid},
and add the frequency dependence based on the Drude formula, 
\begin{equation}
    \kappa(\omega) =%\frac{1}{V} \sum_{\mathbf{k}} 2\omega_{\mathbf{k}}v_{\mathbf{k},x'}^2\frac{\tau(\mathbf{k})}{1-i\omega_{\mathbf{k}}\tau(\mathbf{k})}\frac{\partial}{\partial T}n(\mathbf{\mathbf{k}}) \label{cvini2},
    \iint \dd\mathbf{k} c_v(\mathbf{k})v_{\mathbf{k},x'}^2\frac{\tau_{r}(\mathbf{k})}{1-i\omega\tau_{r}(\mathbf{k})} \label{spinwavekappa}
\end{equation}
where $v_{\mathbf{k},x'} = \partial \omega_{\mathbf{k}}/\partial k_{x'}$ is the magnon velocity along the $x'$ direction, and $\tau_{r}(\mathbf{k})$ is the $\mathbf{k}$-dependent relaxation time. 
As an analog, one can refer to the photonic thermal conductivity for the Debye model~\cite{PhysRevLett.87.047202,PhysRevB.67.184502}.
The only unknown quantity in Eq.~\ref{spinwavekappa} is $\tau_{r}(\mathbf{k})$. However, we can obtain the Drude weight explicitly without knowing $\tau_{r}(\mathbf{k})$.
Integrating the real part of Eq.~\ref{spinwavekappa} over $\omega$ gives
\begin{equation}
    \int_0^{\infty}\Re \kappa(\omega)\dd \omega = \frac{\pi}{2}
    \iint \dd\mathbf{k} c_v(\mathbf{k})v_{\mathbf{k},x'}^2.
\end{equation}

\section{Simulation details} \label{sec:parameter}
Measurements in the main text are evaluated with an maximum imaginary time discretization $\dd\tau=0.05/t$, although for high temperatures the smallest $L=\beta/\dd\tau=20$.
Measurements are performed on Markov chains, which consist of $5\times 10^4$ warm up steps and $10^6$ measurement steps.
Unequal time measurements are taken every $4$ measurement steps.
For lattice sizes between $4 \times 4$ and $8 \times 8$, we measure up to $\sim 1600$ Markov chains (even more for some parameters) for each set of parameters at the lowest temperatures, leading to $2.5\times 10^5$ unequal time measurements and $2\times 10^5 \times L$ equal time measurements per Markov chain.
For relatively high temperatures, fewer Markov chains are simulated. 
For lattice size $10\times 10$ and $U/t=8$, each Markov chain consists of $2.5\times 10^5$ measurement steps ($6.25\times 10^4$ unequal time measurements and $5\times 10^4 \times L$ equal time measurements); and we measure between $\sim 3200$ and $6400$ Markov chains for each set of parameters.

MaxEnt requires a model function, which at the highest temperature we estimate from the spectrum at infinite temperature using the moments expansion method, as in Ref.~\cite{Huang987}.
To summarize this technique, from Eqs.~\ref{definition1}-\ref{definition4} 
\begin{widetext}
\begin{align}
    & \Re L_{aa}(\omega) \equiv %\Re \frac{1}{\omega V\beta} \int^{\infty}_{-\infty} \dd t e^{i\omega t} \theta (t)\ev{[J_{a,x}(t),J_{a,x}]} \nonumber \\
    \Re \frac{1}{V \beta}\int_0^\infty \dd \tilde{t} e^{i(\omega+i0^+) \tilde{t}}\int_0^\beta\dd\tau\ev{J_{a,x}(\tilde{t}-i\tau)J_{a,x}(0)} \nonumber \\
    &=\frac{-\pi}{V \beta} \sum_{\zeta,\eta} \frac{e^{-\beta E_\zeta}(1- e^{-\beta (E_\eta-E_\zeta)})}{E_\zeta-E_\eta}||\ev{\zeta|J_{a,x}|\eta}||^2 \delta(\omega+E_\zeta-E_\eta)\label{whyeven} \\
    &=\frac{1-e^{-\beta \omega}}{\omega V\beta} \Re \int^{\infty}_0 \dd \tilde{t} e^{i(\omega+i0^+) \tilde{t}} \ev{J_{a,x}(\tilde{t})J_{a,x}} =\frac{1-e^{-\beta \omega}}{2\omega V\beta}  \int^{\infty}_{-\infty} \dd \tilde{t} e^{i\omega \tilde{t}} \ev{J_{a,x}(\tilde{t})J_{a,x}}. \label{equationLaa}
\end{align}
\end{widetext}
Here we have used
\begin{align}
(e^{i(\omega+i0^+) \tilde{t}} \ev{J_{a,x}(\tilde{t})J_{a,x}})^* = e^{-i(\omega-i0^+) \tilde{t}} \ev{J_{a,x}(-\tilde{t})J_{a,x}}.
\end{align}
As $T \rightarrow \infty$ ($\beta \rightarrow 0$), Eq.~\ref{equationLaa} becomes
\begin{align}
    \Re L_{aa}(\omega)  =\frac{1}{2V}  \int^{\infty}_{-\infty} \dd \tilde{t} e^{i\omega \tilde{t}} \ev{J_{a,x}(\tilde{t})J_{a,x}},
\end{align}
or
\begin{align}
    \ev{J_{a,x}(\tilde{t})J_{a,x}} = \frac{V}{\pi}\int^{\infty}_{-\infty}\dd \omega  e^{-i\omega \tilde{t}}  \Re L_{aa}(\omega),
\end{align}
which means $\ev{J_{a,x}(\tilde{t})J_{a,x}}$ is real and even since one can prove $\Re L_{aa}(\omega)$ to be even by interchanging $\zeta$ and $\eta$ in Eq.~\ref{whyeven}.
Finite even order derivatives of $\ev{J_{a,x}(\tilde{t})J_{a,x}}$ at $\tilde{t}=0$ can be calculated analytically at $T=+\infty$,
\begin{align}
   & \frac{\dd^{2l}}{\dd \tilde{t}^{2l} } \ev{J_{a,x}(\tilde{t})J_{a,x}} |_{\tilde{t}=0}= (-1)^{l}\ev{(\mathcal{L}^{2l}J_{a,x})J_{a,x}} \nonumber \\
   & =  \ev{(\mathcal{L}^{l}J_{a,x})( \mathcal{L}^{l} J_{a,x})} ,
\end{align}
which determine the Taylor expansion coefficients of $\ev{J_{a,x}(\tilde{t})J_{a,x}}$.
Here $\mathcal{L}$ is the Liouvillian operator $\mathcal{L}O=[H,O]$ for any operator $O$ and we have used $\ev{(\mathcal{L}O_1)O_2}= -\ev{O_1(\mathcal{L}O_2)}$.
The highest order that we calculate is $l=8$.
We can use these coefficients to calculate the Pad\'{e} approximate of $\ev{J_{a,x}(\tilde{t})J_{a,x}}$.
Within the range of $\tilde{t}$ where Pad\'{e} approximates converge (that is, the expansions at the two highest orders do not deviate from each other within this $\tilde{t}$ range), we fit the highest order approximate to
\begin{align}
    \ev{J_{a,x}(\tilde{t})J_{a,x}} = A_1 \sech(\Gamma_1 \tilde{t}) + A_2 \sech(\Gamma_2 \tilde{t}) \cos (\omega_2 \tilde{t}),
\end{align}
where $A_1$, $A_2$, $\Gamma_1$, $\Gamma_2$ and $\omega_2$ are the fitting parameters. Therefore we have an estimate of $\ev{J_{a,x}(\tilde{t})J_{a,x}}$ to convert to $\Re L_{aa}(\omega)$ for the model function.
We note as before that the coefficients $\Re L_{aa}(\omega)$ are positive definite in Eq.~\ref{whyeven}, allowing us to use MaxEnt analytic continuation~\cite{ana1,ana2} to obtain the coefficients in real frequency.  To determine the adjustable parameter in MaxEnt, we use the method in Ref.~\cite{PhysRevE.94.023303}. We use an annealing procedure to determine the model function for other temperatures, which means we use the spectra from the current temperature as the model function for the next lower temperature.
This strategy is the same as that used in Ref.~\cite{Huang987}.  
For lattice sizes other than $8\times 8$, where data is only obtained in the low-temperature regime, the spectrum at infinite temperature is no longer a good approximation for the initial model function.
In such cases, we use the spectrum at a higher temperature, obtained from the simulations on $8\times 8$ lattices, for the initial model function.
To determine the error for results from analytic continuation, we calculate $200$ bootstraps~\cite{bootstrap} (with the same model function) and calculate the standard error from resampling.

\section{Effects of non-zero $t'$} \label{sec:nnn}

Next-nearest-neighbour hoping $t'$ is usually non-zero in cuprates but will lead to a sign problem in our model even at half-filling~\cite{SIGN}. However, we still can access temperatures lower than $J$. 

\begin{figure}
    \centering
    \includegraphics[width=\textwidth]{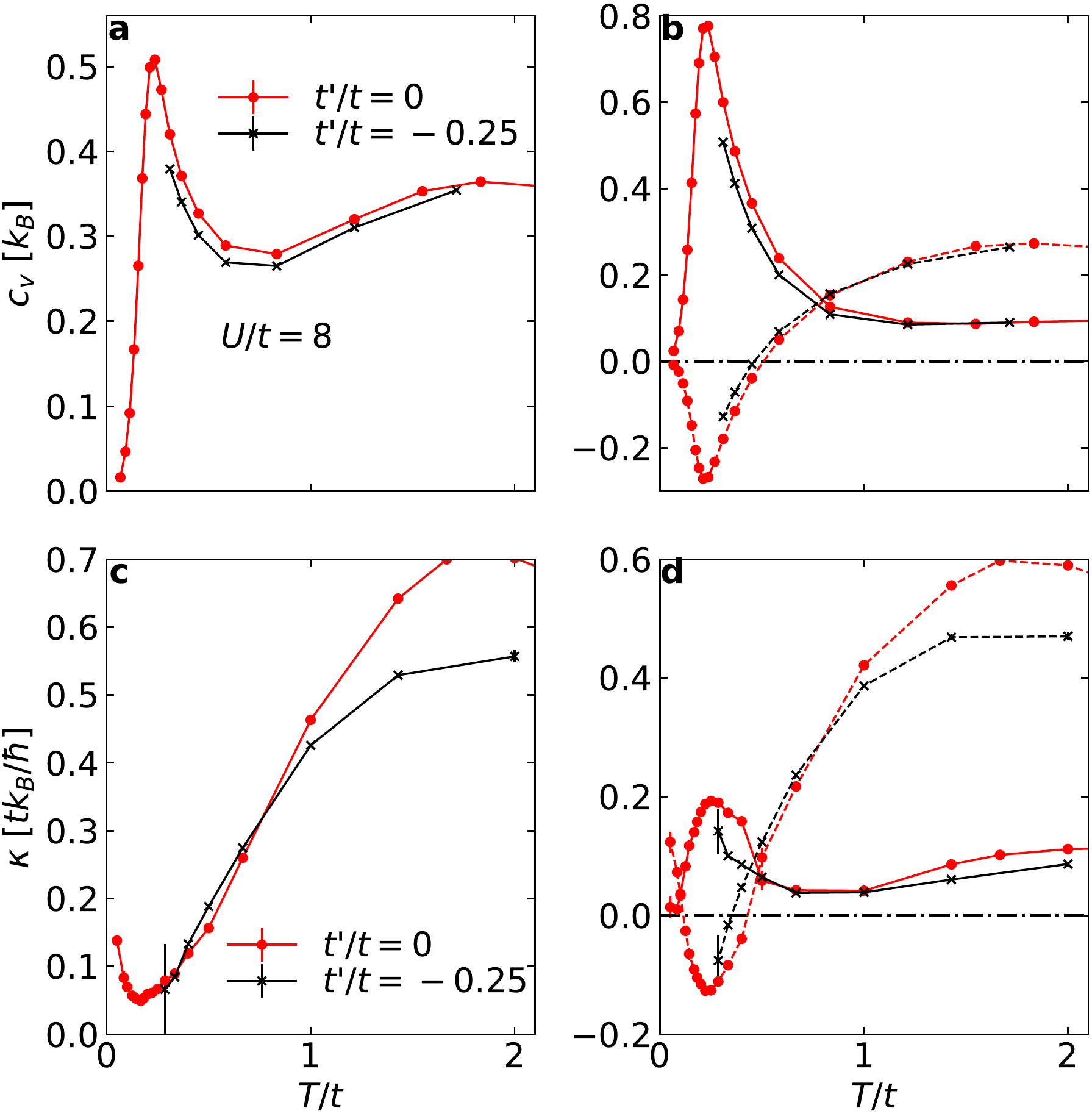}
    \caption{Comparison of specific heat $c_v$ and DC thermal conductivity $\kappa$ between $t'=0$ (the same data as Fig.~2 in the main text) and $t'/t=-0.25$ at half-filling for $U/t=8$.
    (\textbf{a}) The total specific heat $c_v$ calculated by finite differences.
    (\textbf{b}) The kinetic part $c_K$ (solid lines) and potential part $c_P$ (dashed lines)  of the specific heat $c_v$ 
    calculated by finite differences.
    (\textbf{c}) The total DC thermal conductivity $\kappa$.
    (\textbf{d}) The kinetic part $\kappa_K$ (solid lines) and potential part $\kappa_P$ (dashed lines)  of $\kappa$.
Calculation of error bars is the same as Fig.~2 in the main text.    
Simulation lattice size is $8 \times 8$.
Imaginary time discretization $\dd\tau=1/(20t)$.}
    \label{fig:tpsummary}
\end{figure}

In Fig.~\ref{fig:tpsummary}, we compare $c_v$ and $\kappa$, as well as their kinetic-potential separation between next-nearest-neighbour hopping $t'/t=0$ and $t'/t=-0.25$ for $U/t=8$ at half-filling.
We observe that the qualitative behavior is not affected by $t'$, 
down to the lowest temperatures.

Here we introduce the required changes to our methodology for non-zero $t'$.
The kinetic energy current $\mathbf{J}_K$ now includes terms $\propto t^2, tt'$ and $t'^2$, and the potential energy current $\mathbf{J}_P$ includes terms $\propto Ut$ and $Ut'$.  Since we do not have particle-hole symmetry for $t'/t=-0.25$, the chemical potential can be non-zero. 
For convenience, we include the additional term $-\mu\mathbf{J}$ in $\mathbf{J}_P$, such that $\mathbf{J}_Q=\mathbf{J}_K+\mathbf{J}_P$. 
The value of neither $\kappa_P$ nor $\kappa_K$ is affected by this addition as we demonstrated in Sec.~\ref{formula}. 
$\chi_{H_KN}$, $\chi_{H_PN}$, $\ev{\mathbf{J}_K\mathbf{J}}$, and $\ev{\mathbf{J}_P\mathbf{J}}$ also can be non-zero without particle-hole symmetry.
Therefore, for the thermal conductivity we analytically continue $\ev{T_\tau O(\tau)O(0)}$, where $O=J_{x}$, $J_{Q,x}$, $J_{K,x}$, $J_{P,x}$, $J_{Q,x}+J_{x}$,  $J_{K,x}+J_{x}$, and $J_{P,x}+J_{x}$, to calculate the corresponding DC coefficients, and obtain $\kappa_K$, $\kappa_P$, and $\kappa$ by appropriate combinations of these coefficients.
In the MaxEnt analytic continuation, we use a flat model function for all temperatures.
To determine the error, we calculate $200$ bootstraps and calculate the standard error from resampling, as we do for $t'=0$.
\bibliography{main}